\documentclass[letterpaper, 10 pt, conference]{ieeeconf} 
\IEEEoverridecommandlockouts
\usepackage{amsmath,amssymb}
\usepackage{graphicx}
\usepackage{bm}
\usepackage{textcomp}
\usepackage{xcolor}
\usepackage{algorithm}
\usepackage{bbm}
\usepackage{cite}
\usepackage{arydshln}  
\usepackage{xprintlen}
\usepackage{cases}
\usepackage{tabularx,ragged2e}
\usepackage{booktabs}
\usepackage{nccmath}
\usepackage{caption}
\usepackage{float}
\usepackage{subcaption}
\usepackage{mathtools}
\usepackage{algorithm}
\usepackage{algpseudocode}
\usepackage{svg}
\def\BibTeX{{\rm B\kern-.05em{\sc i\kern-.025em b}\kern-.08em
    T\kern-.1667em\lower.7ex\hbox{E}\kern-.125emX}}

\renewcommand{\vec}[1]{\bm{#1}}
\newcommand{\true}{\top}
\newcommand{\false}{\bot}

\newtheorem{problem}{\textbf{Problem}}
\newtheorem{proposition}{\bf{Proposition}}
\newtheorem{definition}{\bf{Definition}}
\newtheorem{lemma}{\bf{Lemma}}
\newtheorem{theorem}{\bf{Theorem}}
\newtheorem{remark}{\bf{Remark}}
\newtheorem{assumption}{\bf{Assumption}}

\newtheorem{example}{\textbf{Example}}
\newtheorem{Fact}{\textbf{Fact}}

\newcommand{\comp}[2]{
\left[
#1 \mid #2 \vphantom{X}
\right]
}
\begin{document}

\title{A Communication Consistent Approach to Signal Temporal Logic Task Decomposition in Multi-Agent Systems   }

\author{Gregorio Marchesini, Siyuan Liu, 
Lars Lindemann and Dimos V. Dimarogonas
	\thanks{This work was supported in part by the Horizon Europe EIC project SymAware (101070802), 
the ERC LEAFHOUND Project, the Swedish Research Council (VR), Digital Futures, and the Knut and Alice Wallenberg (KAW) Foundation.}
\thanks{Gregorio Marchesini, Siyuan Liu, and Dimos V. Dimarogonas are with the Division
of Decision and Control Systems, KTH Royal Institute of Technology, Stockholm, Sweden.
	E-mail: {\tt\small \{gremar,siyliu,dimos\}@kth.se}.  Lars Lindemann is with the Thomas Lord Department of Computer Science, University of Southern California, Los Angeles, CA, USA.
	E-mail: {\tt\small \{llindema\}@usc.ed}.   
}
}

\maketitle
\begin{abstract}
We consider the problem of decomposing a global task assigned to a multi-agent system, expressed as a formula within a fragment of Signal Temporal Logic (STL), under range-limited communication. Given a global task expressed as a conjunction of local tasks defined over the individual and relative states of agents in the system, we propose representing task dependencies among agents as edges of a suitably defined task graph. At the same time, range-limited communication naturally induces the definition of a communication graph that defines which agents have access to each other's states. Within these settings, inconsistencies arise when a task dependency between a pair of agents is not supported by a corresponding communication link due to the limited communication range. As a result, state feedback control laws previously derived to achieve the tasks' satisfaction can not be leveraged. We propose a task decomposition mechanism to distribute tasks assigned to pairs of non-communicating agents in the system as conjunctions of tasks defined over the relative states of communicating agents, thus enforcing consistency between task and communication graphs. Assuming the super-level sets of the predicate functions composing the STL tasks are bounded polytopes, our task decomposition mechanism can be cast as a parameter optimization problem and solved via state-of-the-art decentralized convex optimization algorithms. To guarantee the soundness of our approach, we present various conditions under which the tasks defined in the applied STL fragment are unsatisfiable, and we show sufficient conditions such that our decomposition approach yields satisfiable global tasks after decomposition.
\end{abstract}
\section{Introduction}
Recently, Signal Temporal Logic (STL) has received increasing attention as a tool to monitor and control the run-time behaviour of multi-agent cyber-physical systems \cite{maler2004monitoring}. Notably, an extensive body of literature has been developed dealing with synthesizing admissible plans and feedback solutions that achieve the satisfaction of a given global STL task assigned to a multi-agent system \cite{DonzeAndRaman,farahani2015robust,adrian,MILP4,sun2022multi,ppcLars,eventstlplanning,decenralisedAssumeGuaranteeMILP,siyuanWork,MultiAgentSTL1}. While the scalability of the proposed approaches varies depending on the semantic richness of the considered task and the number of agents involved in the task, task decomposition has been recently explored as a solution to mitigate the curse of dimensionality at the control and planning level, borrowing from similar approaches applied for other types of temporal logics \cite{luo2022temporal,schillinger2018simultaneous}. Namely, the authors in \cite{leahy2022fast,leahy2023rewrite,wang2023controller,charitidou2021signal} first proposed optimization-based approaches to split complex global STL tasks into simpler ones assigned to teams of agents in the systems, whose satisfaction, when jointly considered, recovers the satisfaction of the original global task. However, the influence of range-limited communication has received only minor attention in these settings \cite{liu1,xu2019controller,buyukkoccak2021distributed,marchesini2024communication}. Indeed, in environments where all-to-all communication can not be ensured, consistency between task dependencies and communication is the key factor enabling many of the previous control/planning approaches to be reactive to changes in the environment. From this observation, the contribution of this work is to establish an approach whereby agents can, in a decentralized fashion, reassign task dependencies that are not consistent with the range-limited communication constraints while recovering the original task satisfaction.

We consider the setting in which the global STL task is expressed as a conjunction of local tasks assigned to single agents and pairs of agents in the system. These kinds of specifications are particularly suited for time-varying formation control of multi-agent systems, which find wide applications in robotics in the context of surveillance and
reconnaissance tasks, localization and exploration tasks, and coverage control, to mention a few \cite{Dong,alonso2019distributed,yang2021survey,rahimi2014time}. Within this framework, task dependencies among pairs of agents are captured by a task graph. At the same time, range-limited communication naturally induces the definition of a communication graph where edges are defined by pairs of communicating agents. We propose to decompose tasks assigned over pairs of non-communicating agents as a conjunction of tasks defined over agents forming a multi-hop communication path between the former. More specifically, we show that by considering concave predicate functions whose super-level set is defined by a bounded polytope, our decomposition mechanism can be cast as a convex optimization problem. In addition, we show that such optimization can be efficiently decentralized over the edges of the communication graph when its structure is forced to be acyclic by letting each agent select a limited number of communication links, thus providing scalability of the proposed approach. Eventually, we provide sufficient conditions under which the new task derived from the decomposition is satisfiable and its satisfaction implies the original global task. To the best of our knowledge, this work represents the first decentralized solution to an STL task decomposition problem in multi-agent settings.\par
Previous literature in STL task decomposition includes the work by \cite{charitidou2021signal}, which first introduces a continuous-time and continuous-space STL task decomposition via convex optimization, which we build upon. On the other hand, the authors in \cite{leahy2022fast, leahy2023rewrite} employ a Mixed-integer Linear Program (MILP)s formulation to decompose tasks expressed in a variant of the STL formalism, denoted as Capability STL (CaSTL), as local team-wise tasks. Differently from these previous works, our decomposition approach considers communication constraints at the task decomposition level. It is noted however that the STL fragment considered in \cite{leahy2022fast, leahy2023rewrite} is syntactically richer than the one considered in \cite{charitidou2021signal} and our work. The relevance of this work should be considered with respect to previous reactive feedback control solutions for the satisfaction of STL tasks in multi-agent settings, where all-to-all agent communication is explicitly or implicitly assumed  \cite{LarsControl2,charitidou2021barrier,Gundana,xu2019controller}. Namely, we aim to mitigate the full connectivity requirement for this previous literature by decomposing tasks defined over non-communicating agents into conjunctions of tasks over those that do communicate through the communication network. Since decomposition occurs offline, our method retains the reactive, adaptive qualities of previous feedback control laws in range-limited communication settings. We list the following contributions:\par
i) We provide a task decomposition mechanism to enforce consistency between the task and communication dependencies, enabling previously developed feedback control laws to be leveraged for the satisfaction of a global STL task.\par
ii) We prove that by assuming the communication graph to be acyclic,  our decomposition can be decentralized over the edges of the communication graph and solved by state-of-the-art decentralized convex optimization algorithms as the ones proposed in \cite{notarnicola2019constraint,falsone2017dual,boyd2011distributed}.\par
iii) We provide a set of conditions under which conjunctions of collaborative tasks over the proposed task graph are unsatisfiable (conflicting conjunctions hereafter).\par
iv) We derive sufficient conditions under which our decomposition approach of the original global task assigned to the system yields a new global task that does not suffer from the aforementioned types of conflicting conjunctions and such that its satisfaction implies the original one.\par
The rest of the paper is organized as follows: Section \ref{Preliminaries} reviews polytope operations, STL, and graph theory. The problem statement is provided in Section \ref{problemformulation}. Section \ref{conflicting conjunctions subsection} introduces conflicting conjunctions in collaborative tasks, while Sections \ref{task decomposition}-\ref{optimization section} develop the task decomposition approach. Numerical simulations and conclusions are presented in Sections \ref{simulations} and \ref{conclusions}.

\section{Notation and Preliminaries}\label{Preliminaries}
Bold letters denote vectors, capital letters indicate matrices and sets. Vectors are considered to be columns. Given $\vec{a}_i\in \mathbb{R}^{n_i}$, the vector $\vec{a}  = [\vec{a}_i]_{i\in \mathcal{I}} \in \mathbb{R}^{\sum_i n_i}$ is obtained by vertically stacking the vectors $\vec{a}_i$ with index $i\in \mathcal{I}$. For a matrix $A\in \mathbb{R}^{n\times n}$ and vector $\vec{a}\in \mathbb{R}^n$, the matrix $\comp{A}{\vec{a}} \in \mathbb{R}^{n\times (n+1)}$ is obtained by stacking $A$ and $\vec{a}$ horizontally. The notation $\vec{x}[k]$ indicates the \textit{k}$^{th}$ element of $\vec{x}$ and $A[i,j]$ indicates the element at the \textit{i}$^{th}$ row and \textit{j}$^{th}$ column of the matrix $A$. Given a vector $\vec{x}\in \mathbb{R}^n$, the standard notation for the 2-norm $\|\vec{x}\| = \sqrt{\sum_{k} \vec{x}[k]^2}$ applies, while we adopt the notation  $\langle\vec{x}\rangle = \min_k \{\vec{x}[k]\}$ to indicate the element-wise minimum. The notations $|\mathcal{A}|$ and $2^\mathcal{A}$ denote cardinality and the power set of $\mathcal{A}$, respectively,  while the symbols $\oplus$, $\ominus$ and $\bigtimes$ indicate the Minkowski sum, Minkowski difference and Cartesian product among sets, respectively. The symbol $\otimes$ indicates the Kronecker product of two matrices/vectors. Let $blk(A_1,\ldots A_N)$ indicate the block diagonal matrix with blocks $A_1,\ldots A_N$. The notation $\vec{1}_{m\times n}$, $\vec{1}_{n}$ , $\vec{0}_{m\times n}$ and  $\vec{0}_{n}$ indicate matrices/vectors of ones and zeros respectively while $I_n$ indicates the $n$-dimensional identity matrix. The set $\mathbb{R}_{+}$ denotes the non-negative real numbers.


Let $\mathcal{V}=\{1,\ldots N\}$ be the set of indices assigned to each agent in a multi-agent system and let each agent be governed by the input-affine nonlinear dynamics:
\begin{equation}\label{eq:single agent dynamics}
\dot{\vec{x}}_i = f_i(\vec{x}_i) + g_i(\vec{x}_i)\vec{u}_i, \; 
\end{equation}
where $\vec{x}_i \in \mathbb{X}_i \subset \mathbb{R}^n$ and  $\vec{u}_i\in \mathbb{U}_i\subset \mathbb{R}^{m_i}$ represent the state and control input for agent $i$ with input dimensions $m_i\geq1$. Without loss of generality,  consider $\mathbb{X}_i$ and $\mathbb{U}_i$ to be compact sets containing the origin. Let $f_i : \mathbb{X}_i \rightarrow  \mathbb{R}^n$, $g_i : \mathbb{X}_i \rightarrow \mathbb{R}^{n\times m_i}$ be locally Lipschitz continuous functions on $\mathbb{X}_i$.
Furthermore, let $\vec{u}_i : \mathbb{R}_+ \rightarrow \mathbb{U}_i \in \mathcal{U}_i$ and $\vec{x}_i : \mathbb{R}_+ \rightarrow \mathbb{X}_i \in \mathcal{X}_i $ represent the input and state \textit{signals} for each agent $i$ such that $\mathcal{U}_i$ is the set of Lipschitz continuous input signals and $\mathcal{X}_i$ is the set of absolutely continuous solutions of \eqref{eq:single agent dynamics} under the input signals from $\mathcal{U}_i$.
The global MAS dynamics is then compactly written as
\begin{equation}\label{eq:multi agent dynamics}
    \dot{\vec{x}} = f(\vec{x}) + g(\vec{x})\vec{u},
\end{equation}
with $\vec{x}:= [\vec{x}_i]_{i\in \mathcal{V}} \in \mathbb{X}, \vec{u}:= [\vec{u}_i]_{i\in \mathcal{V}} \in \mathbb{U}, f(\vec{x}):=[f_i(\vec{x}_i)]_{i\in \mathcal{V}}$, $g(\vec{x}):=blk(g_1(\vec{x}_1),\ldots g_N(\vec{x}_N))$,   $\mathbb{X}:=\bigtimes_{i\in \mathcal{V}} \mathbb{X}_i$ and $\mathbb{U}:=\bigtimes_{i\in \mathcal{V}} \mathbb{U}_i$. Consistently, the definition of state and input signals for the MAS are given as $\vec{x}(t) = [\vec{x}_i(t)]_{i\in \mathcal{V}} \in \mathcal{X}$ and $ \vec{u}:= [\vec{u}_i(t)]_{i\in \mathcal{V}} \in \mathcal{U}$ with $\mathcal{X} := \bigtimes_{i \in \mathcal{V}} \mathcal{X}_i$ and $\mathcal{U} := \bigtimes_{i \in \mathcal{V}} \mathcal{U}_i$. Let $\vec{e}_{ij}:=\vec{x}_j-\vec{x}_{i} \in \mathbb{X}_{ij}$ represent the relative state vector for each $i,j\in\mathcal{V}$ where $\mathbb{X}_{ij}:= \mathbb{X}_j \ominus \mathbb{X}_i$ and let $S\in \mathbb{R}^{n_p\times n}$ be a selection matrix\footnote{A selection matrix applies to select $m\leq n$ unique elements from a vector of dimensions $n$ such that $S\in \mathbb{R}^{m \times n}$ with $S[i,j] \in \{0,1\}$;  $\sum_{j=1}^{n}S[i,j] =1, \, \forall i=1,\ldots m$ and $\sum_{i=1}^{m}S[i,j] =1, \, \forall j=1,\ldots n$.} such that $\vec{p}_i = S\vec{x}_i, \; \forall i\in \mathcal{V}$, represents the position vector of agent $i$ with dimension $n_p\leq n$.

\subsection{Polytopes and their properties}
In this subsection, some fundamental properties of polytopes based on \cite[Ch. 0-1]{ziegler2012lectures} and \cite[Sec. 3]{rockafellar1997convex} are revised and adapted to the settings of this work for clarity of presentation.
\begin{definition}{(\cite[pp. 28]{ziegler2012lectures})}\label{h-polyedron}
Given  $A\in \mathbb{R}^{m\times n}$, $\vec{z}\in \mathbb{R}^{m}$ and a center $\vec{c}\in \mathbb{R}^{n}$, the set 
$\mathcal{P}(A,\vec{c},\vec{z}):=\{\vec{x}\in\mathbb{R}^n\, |\,A(\vec{x}-\vec{c})-\vec{z} \leq \vec{0} \} \subset \mathbb{R}^n$  is a \textit{polytope} if it is bounded.
\end{definition}

Note that the inequality in the definition of $\mathcal{P}(A,\vec{c},\vec{z})$ should be interpreted row-wise. Moreover, the centre vector $\vec{c}$ is here introduced for the convenience of analysis, but the polytope definition $\mathcal{P}(A,\tilde{z})=\{\vec{x}\in \mathbb{R}^n | A\vec{x}\leq \tilde{\vec{z}}\}$ in \cite[pp. 28]{ziegler2012lectures} is equivalent to ours letting $\tilde{\vec{z}} = \vec{z} + A\vec{c}$.
\begin{definition}\label{convex hull}(\cite[pp. 4]{ziegler2012lectures})
Given a finite set of points $V = \{\vec{v}_k\}_{k=1}^{|V|} \subset \mathbb{R}^n$ with $|V|\geq1$, the \textit{convex hull} of $V$ is defined as $conv(V):=\{\sum_{k=1}^{|V|}\lambda_k\vec{v}_k | \sum_{k=1}^{|V|} \lambda_k=1,\,\lambda_k\geq0,\, \forall k= 1,\ldots |V| \}$.
\end{definition}
\begin{proposition}\label{equivalent represenations}
(\cite[Thm. 1.1]{ziegler2012lectures})
The set $\mathcal{P}(A,\vec{c},\vec{z})$ is a polytope if and only if there exists a finite \textit{generator} set $V$ such that $\mathcal{P}(A,\vec{c},\vec{z}) = conv(V) \subset \mathbb{R}^n$. Furthermore, $\mathcal{P}(A,\vec{c},\vec{z}) = \{\vec{x}\in \mathbb{R}^n|  h(\vec{x}) \geq 0 \}$ where $h(\vec{x}):= \langle -( A(\vec{x} - \vec{c})-\vec{z})\rangle$.
\end{proposition}

Proposition \ref{equivalent represenations} establishes three equivalent representations for a polytope $\mathcal{P}(A,\vec{c},\vec{z})$. Given  $A$, $\vec{z}$ and $\vec{c}$, then a generator set $V$ can be obtained algorithmically (see for example \cite{avis2000revised}), while the functional representation $h(\vec{x})$ is obtained at no additional computational cost \cite{ziegler2012lectures}. While for a given polytope the generator set $V$ is not unique, we define  $\nu : \mathcal{P}(A,\vec{c},\vec{z}) \mapsto V^*$ as the map between a polytope  $\mathcal{P}(A,\vec{c},\vec{z})$ and the set of unique generators with minimal cardinality $V^*$ (the set of vertices for $\mathcal{P}$ \cite[Prop. 2.2]{ziegler2012lectures}). We next define when two polytopes are said to be \textit{similar} and what properties can be derived from this similarity relation.
\begin{definition}\label{similar}
    Given a polytope $\mathcal{P}(A,\vec{c},\vec{z})$, a scale factor $\alpha>0$ and a center $\bar{\vec{c}}\in \mathbb{R}^n$, then the polytope $\mathcal{P}(A,\vec{c}+\bar{\vec{c}},\alpha\vec{z})$ is \textit{similar} to $\mathcal{P}(A,\vec{c},\vec{z})$.
\end{definition}

Thus two polytopes are similar if one is obtained from the other by homogeneous scaling and a translation. Throughout the presentation, the column vector $\vec{\eta} = [\vec{c},\alpha ] \in \mathbb{R}^{n+1}$ compactly represents this similarity transformation. The following proposition relates the generator sets of two similar polytopes.
\begin{proposition}\label{linear scaling and translation}
Let the polytope $\mathcal{P}(\vec{A},0,\vec{z})$, the center vector $\vec{c}\in \mathbb{R}^n$ and scale $\alpha > 0$ such that $\vec{\eta} = [\vec{c},\alpha ] \in \mathbb{R}^{n+1}$. Moreover, let $V = \nu(\mathcal{P}(A,\vec{0},\vec{z}))$ be the generator set, such that $V=\{\vec{v}_k\}_{k=1}^{|V|}$, then 
\begin{equation}\label{eq:shifted generators}
\nu(\mathcal{P}(A,\vec{c},\alpha\vec{z}))=    \{G_k \vec{\eta}\}_{k=1}^{|V|}, 
\end{equation}
where $G_k = \comp{I_n}{\vec{v}_k} \in \mathbb{R}^{n\times (n+1)}, \vec{v}_k \in V, \forall k=1,\ldots |V|$.
\end{proposition}
\begin{proof}
    Given in Appendix \ref{appendixa}.
\end{proof}

\begin{figure}
    \centering
    \includegraphics[width=0.5\textwidth]{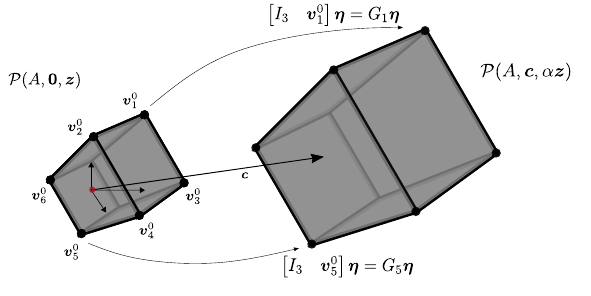}
    \caption{Graphical representation of similar polytopes. On the left, the polytope $\mathcal{P}(\vec{A},\vec{0},\vec{z})$ with a red dot representing the origin of a common reference system. On the right, the scaled and shifted polytope $\mathcal{P}(\vec{A},\vec{c},\alpha \vec{z})$.}
    \label{fig:shifted cubes}
\end{figure}
In other words, the generators of a polytope $\mathcal{P}(A,\vec{c},\alpha\vec{z})$ are directly computed by translating and scaling the generators of the similar polytope $\mathcal{P}(A,\vec{0},\vec{z})$ as represented in Fig. \ref{fig:shifted cubes}. The next proposition extends the result of Prop. \ref{linear scaling and translation} to the Minkowski sum of similar polytopes.
\begin{proposition}\label{sequence of polytopes}
Let  $\mathcal{P}(A,\vec{c}_i,\alpha_i\vec{z}), \, \forall i\in\mathcal{I}$ be similar polytopes for some index set $\mathcal{I}$ and $\alpha_i\in \mathbb{R}_{+}$, then
\begin{equation}\label{eq:set summation}
\bigoplus_{i\in \mathcal{I}}\mathcal{P}(A,\vec{c}_i,\alpha_i\vec{z}) = \mathcal{P}(A,\sum_{i\in \mathcal{I}}\vec{c}_i,\sum_{i\in \mathcal{I}}\alpha_i\vec{z}).
\end{equation}
Moreover, let $\vec{\eta}_i = [\vec{c}_i,\alpha_i]$ and the generator set $V= \nu(\mathcal{P}(A,\vec{0}, \vec{z}))=\{\vec{v}_k\}_{k=1}^{|V|}$, then it holds 
\begin{equation} \label{eq:generators equivalence}
\nu\Bigl(\bigoplus\limits_{i\in \mathcal{I}}\mathcal{P}(A,\vec{c}_i,\alpha_i\vec{z})\Bigr)= \{\sum_{i\in \mathcal{I}}  G_k \vec{\eta}_i  \}_{k=1}^{|V|},
\end{equation}
with $G_k = \comp{I_n}{\vec{v}_k} \in \mathbb{R}^{n\times (n+1)}, \vec{v}_k \in V$.
\end{proposition}
\begin{proof}
    Follows from \cite[Thm. 3.2]{rockafellar1997convex} and Prop. \ref{linear scaling and translation}.
\end{proof}\par
 Next, Prop. \ref{polytopes inclusion} provides a set of linear inequalities that can be employed to check the intersection and inclusion of two polytopes.
\begin{proposition}\label{polytopes inclusion}
Let the polytopes $\mathcal{P}(A_1,\vec{c}_1,\alpha_1\vec{z}_1)$, $\mathcal{P}(A_2,\vec{c}_2,\alpha_2\vec{z}_2)$ such that $A_1\in \mathbb{R}^{m_1\times n},A_2\in \mathbb{R}^{m_2\times n}$ and let $\vec{\eta}_1 = [ \vec{c}_1, \alpha_1 ]$ and  $\vec{\eta}_2 = [ \vec{c}_2, \alpha_2 ]$. Furthermore, let $V_1 = \nu(\mathcal{P}(A_1,\vec{0},\vec{z}_1)= \{\vec{v}_k\}_{k=1}^{|V_1|}$. Then, the inclusion relation $\mathcal{P}(A_1,\vec{c}_1,\alpha_1\vec{z}_1) \subseteq \mathcal{P}(A_2,\vec{c}_2,\alpha_2\vec{z}_2)$ holds if and only if
\begin{equation}\label{eq:compact inclusion}
M_2 \vec{\eta}_1 - Z_2 \vec{\eta}_2 \leq \vec{0},
\end{equation}
where  
\begin{equation}\label{eq:intersection matrices}
M_2 = \begin{bmatrix}A_2 G_1\\ \vdots \\ A_2 G_{|V_1|}\end{bmatrix}, \quad  Z_2 = \vec{1}_{|V_1|} \otimes \comp{A_1}{\vec{z}_2}
\end{equation}
and $G_k = \comp{I_n}{\vec{v}_k} \in \mathbb{R}^{n\times (n+1)}, \vec{v}_k \in V_1$. On the other hand, the intersection relation $\mathcal{P}(A_1,\vec{c}_1,\alpha_1\vec{z}_1) \cap \mathcal{P}(A_2,\vec{c}_2,\alpha_2\vec{z}_2) \neq \emptyset$ 
 holds if and only if 
\begin{equation} \label{eq:compact intersection}
\begin{aligned}
    A_1\vec{\xi} -   \comp{A_1}{z_1} \vec{\eta}_1 \leq \vec{0},\; A_2\vec{\xi} -   \comp{A_2}{z_2} \vec{\eta}_2 \leq \vec{0}, 
\end{aligned}
\end{equation}
for some vector $\vec{\xi}\in \mathbb{R}^n$.
\end{proposition}
\begin{proof}
Given in Appendix \ref{appendixa}.
\end{proof}
\par The section is concluded with the following result,  whose relevance is clarified in Sec. \ref{task decomposition subsection}. 
\begin{proposition}(From \cite[Thm. 3.2]{rockafellar1997convex})\label{inclusion accuracy}
    Let $\mathcal{P}(A,\vec{c},\vec{z})$ and the similar polytopes $\mathcal{P}(A,\vec{c}_i,\alpha_i\vec{z}),\; \forall i\in \mathcal{I}$ with $\alpha_i \in \mathbb{R}_+$ and index set $\mathcal{I}$, then \\
    
    \begin{equation}
    \begin{aligned}
    &i) \; \bigoplus_{i\in \mathcal{I}} \mathcal{P}(A,\vec{c}_i,\alpha_i\vec{z}) = \mathcal{P}(A,\vec{c},\vec{z}) \Leftrightarrow \\ 
    &\hspace{0.5cm}\sum_{i\in \mathcal{I}}\alpha_i = 1\;  \land \; \bigoplus_{i\in \mathcal{I}} \mathcal{P}(A,\vec{c}_i,\alpha_i\vec{z}) \subseteq \mathcal{P}(A,\vec{c},\vec{z}),
     \end{aligned}
    \end{equation}
    and 
    \begin{equation}
    ii)\; \bigoplus_{i\in \mathcal{I}} \mathcal{P}(A,\vec{c}_i,\alpha_i\vec{z}) \subset \mathcal{P}(A,\vec{c},\vec{z}) \Rightarrow \sum_{i\in \mathcal{I}}\alpha_i < 1.
    \end{equation}
\end{proposition}
\par
Thus Prop. \ref{inclusion accuracy} supports the intuition that 
given a polytope and a Minkowski sum of polytopes similar to the former, then the Minkoswki sum is included in the original polytope only if the sum of the scale factors $\alpha_i$ is less than or equal to 1. The relevance of this fact is clarified in Section \ref{task decomposition subsection}.
\subsection{Signal Temporal Logic}
Signal Temporal Logic (STL) is a predicate logic applied to formally define spatio-temporal behaviours (tasks) for state signals $\vec{x}(t) \in \mathcal{X}$. Let $h: \mathcal{D} \rightarrow \mathbb{R}$, with $\mathcal{D}\subseteq \mathbb{X}$ be a predicate function and let the boolean-valued predicate $$\mu(h(\vec{x})):= \begin{cases}
\true &\text{if} \; h(\vec{x})\geq 0 \\
\false &\text{if} \; h(\vec{x})<0.
\end{cases}$$
Then, the recursively defined STL grammar applies to construct complex specifications from a set of elementary predicates as
$$\phi:= \mu(h(\vec{x}))|\neg \phi|\phi_1\land \phi_2| \phi_1\lor \phi_2|F_{[a,b]}\phi|G_{[a,b]}\phi|\phi_1U_{[a,b]}\phi_2,$$
where $G,F$ and $U$ are the temporal \textit{always}, \textit{eventually} and \textit{until} operators with time interval $[a,b] \subseteq \mathbb{R}_+$, while $\land,\lor$ and $\neg$ represent the logical conjunction, disjunction and negation operators. The qualitative semantic rules defining the conditions under which a state signal $\vec{x}(t)$ \textit{satisfies} a task $\phi$ from time $t$ (written as $(\vec{x}(t),t) \models \phi$) are as follows \cite[Ch 2.2]{maler2004monitoring}:

\begin{subequations}\label{eq:explicit semantics}
\begin{align}
\begin{split}
(\vec{x}(t),t) &\models \mu(h(\vec{x}(t))) \Leftrightarrow \mu(h(\vec{x}(t))) = \top \\
\end{split}\\
\begin{split}
(\vec{x}(t),t)  &\models \neg\phi \Leftrightarrow (\vec{x}(t),t) \not\models \phi \\
\end{split}\\
\begin{split}
(\vec{x}(t),t)  &\models \phi_1 \land \phi_2  \Leftrightarrow (\vec{x}(t),t)  \models \phi_1 \land  (\vec{x}(t),t) \models \phi_2 \\
\end{split}\\
\begin{split}
(\vec{x}(t),t)  &\models \phi_1 \lor \phi_2 \Leftrightarrow (\vec{x}(t),t)  \models \phi_1 \lor  (\vec{x}(t),t) \models \phi_2 \\
\end{split}\\
\begin{split}
(\vec{x}(t),t)  &\models F_{[a,b]}\phi \Leftrightarrow \exists \tau\in [a,b] \;  \text{s.t}\; (\vec{x},t+\tau) \models \phi\\\label{eq:eventually old def}
\end{split}\\
\begin{split}
(\vec{x}(t),t)  &\models G_{[a,b]}\phi \Leftrightarrow (\vec{x},t+\tau) \models \phi, \forall \tau \in [a,b] \\
\end{split}\\
\begin{split}
(\vec{x}(t),t)  &\models \phi_1 U_{[a,b]}\phi_2 \Leftrightarrow \exists \tau \in [a,b] \; \text{s.t}\,\\
 &\hspace{0.1cm}(\vec{x},t+\tau) \models \phi_2, \land \forall \tau' \in [a,\tau] ,\;  (\vec{x},t+ \tau')\models \phi_1.\\
\end{split}
\end{align}
\end{subequations}
In this work, we deal with a fragment of STL having application in the definition of time-varying missions for multi-agent systems. Namely, let 
\begin{subequations}\label{eq:working fragment}
\begin{equation}
\varphi_{i}  := F_{[a,b]}\mu_i|G_{[a,b]}\mu_i,
\label{eq:single agent spec temporal}
\end{equation}
\begin{equation}
\varphi_{ij} :=  F_{[a,b]}\mu_{ij}|G_{[a,b]}\mu_{ij}, \label{eq:multi agent spec temporal}\\
\end{equation}
\begin{equation}
\phi_{ij} := \bigwedge^{K_{ij}}_{k=1} \varphi^k_{ij} , \; \phi_{i}  := \bigwedge^{K_i}_{k=1} \varphi^k_{i}, \label{eq:conjunctions of multiple tasks}
\end{equation}
\end{subequations}
for some $K_i,K_{ij}\in \mathbb{N}$, such that the boolean-valued predicates $\mu_i,\mu_{ij}$ take the form
\begin{subequations}\label{eq:predicates standard form}
\begin{equation}
\mu_i :=
\begin{cases}
\true &\text{if} \; h_i(\vec{x}_i)\geq 0 \\
\false &\text{if} \; h_i(\vec{x}_i)<0 ,
\end{cases}
\end{equation}
\begin{equation}
\mu_{ij} := \begin{cases}
\true &\text{if} \; h_{ij}(\vec{e}_{ij})\geq 0 \\
\false &\text{if} \; h_{ij}(\vec{e}_{ij})<0 .
\end{cases}
\end{equation}
\end{subequations}
where 
\begin{subequations}\label{eq:predicates function standard form}
\begin{equation}
h_{i}(\vec{x}_i) =  \langle -(A_{i}(\vec{x}_i-\vec{c}_{i})-\vec{z}_{i}) \rangle,
\end{equation}
\begin{equation}\label{eq:multi agent predicate}
h_{ij}(\vec{e}_{ij})= \langle-(A_{ij}(\vec{e}_{ij}-\vec{c}_{ij})-\vec{z}_{ij}) \rangle.
\end{equation}
\end{subequations}
for some centres $\vec{c}_i, \vec{c}_{ij} \in \mathbb{R}^{n}$. It is assumed that the super-level sets $\mathcal{B}_i,\mathcal{B}_{ij}$ for $h_i(\vec{x}_i)$, $h_{ij}(\vec{e}_{ij})$ are bounded, thus taking the form  
\begin{equation}\label{eq:super level sets}
\begin{aligned}
\mathcal{B}_i &:=\{ \vec{x}_i\in\mathbb{X}_i|\; h_i(\vec{x}_i)\geq0\}=\mathcal{P}(A_{i},\vec{c}_{i},\vec{z}_{i}),\\
\mathcal{B}_{ij} & :=\{ \vec{e}_{ij}\in\mathbb{X}_{ij}|\; h_{ij}(\vec{e}_{ij})\geq0\}=\mathcal{P}(A_{ij},\vec{c}_{ij},\vec{z}_{ij}).
\end{aligned}
\end{equation}
We refer to tasks of type \eqref{eq:single agent spec temporal} and  \eqref{eq:multi agent spec temporal} as \textit{independent} tasks and \textit{collaborative} tasks respectively. Moreover, we refer to $\mathcal{B}_{ij}$ and $\mathcal{B}_{i}$ as the \textit{truth sets} for tasks $\varphi_{ij}$ and $\varphi_i$ respectively since, by definition, these are the sets where $\mu_{ij} = \top$ and $\mu_{i} = \top$. In other words, independent and collaborative tasks enforce the single state and relative state of a pair of agents to lie within a bounded polytope for a time interval specified by the temporal operator included in the task. Note that \eqref{eq:working fragment} is expressive enough to deal with tasks of type $\mu^1 U_{[a,b]}\mu^2$ since $\mu^1 U_{[a,b]}\mu^2 = G_{[a,\tau]}\mu^1 \land F_{[\tau,\tau]}\mu^2$ with $\tau\in [a,b]$. At the same time, recurrent tasks with period $T$ can be encoded in the fragment as conjunctions of the form $\land_{k} G_{[a+kT,b+kT]}\mu$.\par
Concerning the semantics, \textit{quantitative} semantics are applied in this work instead of the qualitative semantics in \eqref{eq:explicit semantics}. Namely, let the \textit{robustness} measure $\rho^{\phi}(\vec{x}(t),t) : \mathcal{X} \times \mathbb{R}_+ \rightarrow \mathbb{R}$ of task $\phi$ for a state signal $\vec{x}(t)$ be defined as 
\begin{subequations}\label{eq:robust semantics}
\begin{align}
\rho^{\mu}(\vec{x}(t),t)&=h(\vec{x}(t)), \label{eq:predicate robust}\\
\rho^{F_{[a, b]} \phi}(\vec{x}(t),t)&=\max _{\tau \in[a, b]}\rho^\phi\left(\vec{x}, t+\tau \right), \label{eq:eventually robust}\\
\rho^{G_{[a, b]} \phi}(\vec{x}(t),t)&= \min _{\tau \in [a, b]} \rho^\phi\left(\vec{x}, t+\tau\right),\label{eq:always robust}\\
\rho^{\phi_1 \wedge \phi_2}(\vec{x}(t),t)&=\min \left(\rho^{\phi_1}(\vec{x}(t),t), \rho^{\phi_2}(\vec{x}(t),t)\right), \label{eq:conjunction robust} 
\end{align}
\end{subequations}
then considering for convenience the initial time $t=0$ we have the satisfaction relation $\rho^{\phi}(\vec{x}(t),0)\geq 0 \; \Leftrightarrow\; (\vec{x}(t),0) \models \phi$ (with double implication holding due to absence of the negation operator in fragment \eqref{eq:working fragment}) \cite[Sec. 2.2]{raman2015reactive}\cite{donze2010robust}. 
\begin{remark}
While generally nonlinear concave predicate functions are commonly considered in the literature, predicate functions with polytope-like truth sets as in \eqref{eq:predicates function standard form} represent a practical design choice. Indeed, let any two predicate functions $h_1(\vec{x}):\mathcal{D} \rightarrow \mathbb{R}$ and $h_2(\vec{x}):\mathcal{D} \rightarrow \mathbb{R}$ such that the truth set $\mathcal{B} = \{ \vec{x} \; | \; h_1(\vec{x}) \geq 0\}= \{ \vec{x} \; | \; h_2(\vec{x}) \geq 0\} $ then $\mu(h_1(\vec{x})) = \mu(h_2(\vec{x})),\, \forall \vec{x} \in \mathcal{D}$.  Hence, different predicate functions with the same truth set are semantically indistinguishable as per the semantics in \eqref{eq:explicit semantics} and \eqref{eq:robust semantics}. Since any smooth convex set $\mathcal{B}$ can be approximated by a polytope $\mathcal{P}(A,\vec{c},\vec{z})$, with the approximation error decreasing in the number of hyperplanes used for the approximation, predicate functions as per \eqref{eq:predicates function standard form} are considered comparable to nonlinear concave predicate functions, with the critical advantage that set operations like Minkowski sums, set inclusions and intersections can be efficiently undertaken \cite{gruber1982approximation}\cite[Ch. 1.10]{gruber1993handbook}.
\end{remark}
\subsection{Communication and task graphs}
Let $\mathcal{G}(\mathcal{V},\mathcal{E})$ represent an undirected graph over the agents in $\mathcal{V}$ with undirected edge set $\mathcal{E}\subseteq \mathcal{V}\times \mathcal{V}$ such that $(i,j)\in \mathcal{E} \Leftrightarrow (j,i)\in \mathcal{E}$. Furthermore, let the neighbour set for agent $i$ be defined as $\mathcal{N}(i)=\{j\, |\, (i,j) \in \mathcal{E} \land i\neq j\}$. In the following, a distinction is made between two types of graphs: the \textit{communication graph} $\mathcal{G}_{c}(\mathcal{V},\mathcal{E}_c)$ and the \textit{task graph} $\mathcal{G}_{\psi}(\mathcal{V},\mathcal{E}_{\psi})$, such that
\begin{subequations}\label{eq:graphs edges}
\begin{align}
\mathcal{E}_{c} &= \{(i,j) | \; \|\vec{p}_{i}-\vec{p}_{j}\| \leq r_c \land q^c_{ij}=\top \} \label{eq:communication edges}\\
\mathcal{E}_{\psi} &= \{(i,j) | \; \exists \phi_{ij} \} \cup \; \{ (i,i) | \; \exists\phi_i  \}\label{eq:task edges},
\end{align}
\end{subequations}
In \eqref{eq:communication edges}, $r_c>0$ is the communication radius and  $q^c_{ij} \in \{\top,\bot\},\; \forall (i,j) \in \mathcal{V}\times \mathcal{V}$ is a boolean \textit{communication token} that is applied to selectively enable communication among agents that are within the communication radius. For consistentency with self-communication, let $q^c_{ii}=\top,\; \forall i\in \mathcal{V}$. On the other hand, note that by \eqref{eq:task edges} an edge $(i,j) \in \mathcal{E}_{\psi}$ exists whether a collaborative task $\phi_{ij}$ as per \eqref{eq:conjunctions of multiple tasks} exists over the relative state $\vec{e}_{ij}$, while self-loops $(i,i) \in \mathcal{E}_{\psi}$ are induced by independent tasks $\phi_i$. Let $\mathcal{N}_c(i)$ and $\mathcal{N}_{\psi}(i)$ be the neighbour sets for the communication and task graphs respectively. When clear from the context, the shorthand notation $\mathcal{G}_c$ and $\mathcal{G}_{\psi}$ applies. In general, the communication graph is time-varying according to our definition since the position of the agents changes over time. However, we assume a supervisory controller enforcing the connectivity for each edge $(i,j) \in \mathcal{E}_c$ exists, such that the graph $\mathcal{G}_c$ is considered static (time-invariant) and determined at the initial time \cite{zavlanos}. The following definition clarifies when a collaborative task is considered consistent with the communication requirements. 
\begin{definition}\label{communication consisistency}(Communication Consistency)
    Given the communication graph $\mathcal{G}_c$, then the collaborative task $\varphi_{ij}$ with associated truth set $\mathcal{B}_{ij}$ as per \eqref{eq:multi agent spec temporal} is \textit{communication consistent}  if $(i,j)\in \mathcal{E}_c$ and it holds 
    \begin{equation}\label{eq:communication consistency equation}
    \mathcal{B}_{ij} \cap \{\vec{e}_{ij}\in \mathbb{X}_{ij}\; |\; \| \vec{p}_j -\vec{p}_j\| \leq r_c \} \neq \emptyset. 
    \end{equation}
    On the other hand, $\varphi_{ij}$ is \textit{communication inconsistent} if it is not consistent. Moreover,  $\mathcal{G}_{\psi}$ is communication consistent if all the collaborative tasks $\varphi_{ij}^k,\; \forall (i,j) \in \mathcal{E}_{\psi} \; k \in$, are.
\end{definition}\par
Communication consistency as per \eqref{communication consisistency} plays a central role in our task decomposition approach. Namely, relation \eqref{eq:communication consistency equation} entails that there exists a nonempty set of relative configuration $\vec{e}_{ij} \in \mathbb{X}_{ij}$ such that the predicate $\mu_{ij}$, associated with task $\varphi_{ij}$, takes the value $\mu_{ij}=\top$ while agent $i$ and $j$ are within the communication radius. This fact thus ensures the satisfiability of $\varphi_{ij}$ subject to the limited-range communication constraint. For brevity, we will often refer to communication consistent tasks simply as \textit{consistent} without loss of clarity. The following assumption about the communication graphs is considered.
\begin{assumption}\label{acyclic communication}
    The communication graph $\mathcal{G}_c$ is connected and the communication tokens $q_{ij}^c$ are chosen such that $\mathcal{G}_c$ is acyclic.
\end{assumption}\par
Assumption \ref{acyclic communication} is an initialization assumption imposing acyclicity of the communication graph $\mathcal{G}_c$. Our previous work in \cite{marchesini2024communication}, dealing with a similar decomposition, considered the milder assumption of connectivity for $\mathcal{G}_c$. The reason why Assumption \ref{acyclic communication} is considered here instead, relates to the fact that the convexity and decentralization property of the task decomposition approach proposed in the work are lost when cycles are present over $\mathcal{G}_c$. Further details justifying this assumption are given in Sec. \ref{handling conflicting conjunctions}. We highlight that the acyclicity of $\mathcal{G}_c$ can be enforced by appropriately selecting the communication tokens $q_{ij}^c$ at time $t=0$ by means of centralized or decentralized algorithms. One such example consists of electing a root agent and expanding a tree that spans all the agents similar to well-known sampling-based planning algorithms over graphs \cite[Ch. 5]{lavalle2006planning}.

\subsection{Paths, cycles and critical tasks}
For a given graph $\mathcal{G}(\mathcal{V},\mathcal{E})$, let the vector $\vec{\pi}_i^j\in \mathcal{V}^l$ represent a \textit{directed} path of length $l$ defined as a vector of non-repeated indices in $\mathcal{V}$ such that $\vec{\pi}_i^j[k]\in \mathcal{V},\; \forall k=1,\ldots l$ ; $\vec{\pi}_i^j[k] \neq \vec{\pi}_i^j[k+1], \;\forall k=1,\ldots l-1$ ; $(\vec{\pi}_i^j[k],\vec{\pi}_i^j[k+1])\in \mathcal{E}$ and $(\vec{\pi}_i^j[1],\vec{\pi}_{i}^j[l])=(i,j)$.  Similarly, let $\vec{\omega}\in \mathcal{V}^l$ represent a cycle such that $\vec{\omega}$ is a path of length $l$ with a single repeated index as $\vec{\omega}[1]=\vec{\omega}[l]$. Moreover, let $\epsilon:\mathcal{V}^l\rightarrow 2^{\mathcal{E}}$ be a set-valued function yielding the edges along the path $\vec{\pi}_i^j$ as $\epsilon(\vec{\pi}_i^j)=\{(\vec{\pi}_i^j[k],\vec{\pi}_i^j[k+1])\; |\; k=1,\ldots l-1\}$, with the same definition extending to cycles $\vec{\omega}$.  From these definitions, the following relations hold 
\begin{subequations}\label{eq:path and cycles relations}
\begin{align}
\vec{e}_{ij}&=\sum_{(r,s)\in \epsilon(\vec{\pi}_{i}^j)}\bm{e}_{rs}\label{eq:edge sequence},\\
\vec{0}  &=\sum_{(r,s)\in \epsilon(\vec{\omega})}\bm{e}_{rs}\label{eq:edge cycle}.
\end{align}
\end{subequations}

\section{Problem Formulation}\label{problemformulation}
Consider the situation in which a set of tasks in the form \eqref{eq:working fragment} is assigned to a multi-agent system such that a task graph $\mathcal{G}_{\psi}$ can be derived as per \eqref{eq:task edges}. The \textit{global} task $\psi$ assigned to the system is compactly written as
\begin{equation}\label{eq:global specification}
\psi := \bigwedge_{i=1}^N  \left( \phi_{i} \land \bigwedge_{j\in \mathcal{N}_\psi(i)}\phi_{ij} \right).
\end{equation}
Global tasks as per \eqref{eq:global specification} are suitable for defining time-varying relative formations from which complex group behaviours for the MAS can be obtained. Previously developed feedback-based control laws to satisfy STL tasks in these settings take the general form $\vec{u}_i(t)=\vec{u}_i(\vec{x}_i(t),\vec{x}_{\mathcal{N}_c(i)}(t),t)$ with $\vec{x}_{\mathcal{N}_c(i)}(t) = [\vec{x}_i(t)]_{i\in \mathcal{N}_c(i)}$ \cite{liu1,liu2,LarsControl2,charitidou2021barrier}. However, a complete\footnote{A graph is complete if every agent is connected to every other agent.} communication graph $\mathcal{G}_c$ is implicitly assumed such that $\vec{x}_{\mathcal{N}_c(i)}(t) = [\vec{x}_i(t)]_{i\in \mathcal{V}}$. More recently in \cite{gregcdc}, we provided a feedback control approach for the satisfaction of global tasks in the form \eqref{eq:global specification}, assuming that the associated task graph and communication graph are consistent as well as acyclicity of the task graph, which we also assume in the current work. This fact motivates the effort for a decomposition approach for rewriting communication inconsistent tasks $\varphi_{ij}$ (cf. Def. \ref{communication consisistency}) as conjunctions of consistent ones, such that the advantages of state feedback can be leveraged to achieve the satisfaction of the task. The problem is formally stated as follows:
\begin{problem}\label{problem1}
Consider the communication graph $\mathcal{G}_c$, task graph $\mathcal{G}_{\psi}$, the global task $\psi$ as per \eqref{eq:global specification} with the associated collaborative tasks $\phi_{ij}=\land_{k=1}^{K_{ij}} \varphi_{ij}^{k},\; \forall (i,j)\in \mathcal{E}_{\psi}$ as per \eqref{eq:conjunctions of multiple tasks}. Develop a decentralized algorithm that outputs a global task $\bar{\psi}$ with task graph $\mathcal{G}_{\bar{\psi}}$ in the form of
\begin{equation}\label{eq:global specification rewritten}
\bar{\psi} := \bigwedge_{i=1}^N \left(\ \phi_{i} \land \bigwedge_{j\in \mathcal{N}_\psi(i)\cap \mathcal{N}_c(i)}\phi_{ij} \wedge \bigwedge_{j\in \mathcal{N}_\psi(i)\setminus\mathcal{N}_c(i)} \bar{\phi}^{ij}\right),
\end{equation}
where
\begin{subequations}\label{eq:path specification}
\begin{align}
\bar{\phi}^{ij}            &:= \bigwedge_{k=1}^{K_{ij}}(\bar{\phi}^{ij})^k, \label{eq:conjunctions of decomposition}\\
(\bar{\phi}^{ij})^k  &:= \bigwedge_{(r,s) \in \epsilon(\vec{\pi}_i^j)} (\bar{\varphi}^{ij}_{rs})^k,\label{eq:single decomposition}
\end{align}
\end{subequations}
for all inconsistent edges $(i,j) \in \mathcal{E}_\psi\setminus \mathcal{E}_c$ and with tasks $(\bar{\varphi}^{ij}_{rs})^k,\;  \forall (r,s) \in \epsilon(\vec{\pi}_i^j), \forall k=1,\ldots K_{ij}$ being communication consistent as per \eqref{eq:multi agent spec temporal}. Moreover, $\bar{\psi}$ should be such that
\begin{equation}\label{eq:task decomposition implication}
(\vec{x}(t),t)\models \bar{\psi} \Rightarrow (\vec{x}(t),t)\models \psi.
\end{equation}
\end{problem}

\begin{figure}[t]
    \centering
    \includegraphics{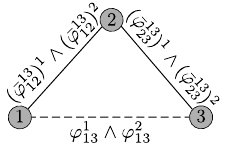}
    \caption{Graphical decomposition of the inconsistent task $\phi_{13}=\varphi_{13}^1 \land \varphi_{13}^2$ over the path $\vec{\pi}_{i}^{j}=[1,2,3]$. Solid edges correspond to edges in the communication graph $\mathcal{G}_c$ and edge $(1,3)$ (dashed) is the only edge in the task graph $\mathcal{G}_{\psi}$.}
    \label{fig:multiple layers conjuncition}
\end{figure}

In other words, by recalling that each inconsistent task is defined as $\phi_{ij}=\land_{k=1}^{K_{ij}}\varphi_{ij}^k$, then the objective in Problem \ref{problem1} is to find a conjunction of task $(\bar{\phi}^{ij})^k$, as per \eqref{eq:single decomposition}, replacing the task $\varphi_{ij}^k$ for all $k=1,\ldots K_{ij}$. Note that each $(\bar{\phi}^{ij})^k$ is defined as a conjunction of new collaborative tasks $(\bar{\varphi}_{rs}^{ij})^k$ over the edges $(r,s)\in \epsilon(\vec{\pi}_i^j) \subseteq \mathcal{E}_c$, where $\vec{\pi}_i^j$ is a path over $\mathcal{G}_c$ connecting $i$ to $j$. The following example provides further intuition on the meaning of the proposed decomposition.
\begin{example}
Consider Figure \ref{fig:multiple layers conjuncition} where two communication edges $(1,2),(2,3) \in \mathcal{E}_c$ (solid edges) are applied to decompose the inconsistent task $\phi_{13}=\varphi_{13}^1 \land \varphi_{13}^2$ (dashed line). The decomposition of $\phi_{13}$ yields the task $\bar{\phi}^{13} = (\bar{\phi}^{13})^1 \land (\bar{\phi}^{13})^2$ where $(\bar{\phi}^{13})^1 = (\bar{\varphi}^{13}_{12})^1 \land (\bar{\varphi}^{13}_{23})^1$ and  $(\bar{\phi}^{13})^2 = (\bar{\varphi}^{13}_{12})^2 \land (\bar{\varphi}^{13}_{23})^2$, as per \eqref{eq:single decomposition}. Hence, the conjunctions $(\bar{\phi}^{13})^1$ and $(\bar{\phi}^{13})^2$ replace $\varphi_{13}^1$ and $\varphi_{13}^2$, respectively.
\end{example} 
\par 
In Section \ref{task decomposition subsection} it is detailed how the tasks $\bar{\phi}^{ij}$ are designed. But before, in the next section, some important results are presented to understand the process under which the tasks $\bar{\phi}^{ij}$ can be introduced over the task graph $\mathcal{G}_{\psi}$ without causing conflicts.
\section{Conflicting conjunctions  over task graphs}\label{conflicting conjunctions subsection}
The concept of conflicting conjunction is introduced first.
\begin{definition}(Conflicting conjunction)\label{conflicting conjunction}
The conjunction of collaborative tasks $\bigwedge_{k=1}^{K_{ij}}\varphi^{k}_{ij}$ is a \textit{conflicting conjunction} if there does not exist a continuous state signal $\vec{x}(t)$ for the multi-agent system such that $(\vec{x}(t),t)\models\bigwedge_{k=1}^{K_{ij}}\varphi^{k}_{ij}$.
\end{definition}
\par 
It is desired that any formula $\bar{\psi}$ as per \eqref{eq:global specification rewritten}, resulting from the decomposition approach, should not induce conflicting conjunctions as this fact would result directly in  $\bar{\psi}$ being unsatisfiable. We here study five types of conflicting conjunctions according to Def. \ref{conflicting conjunction} which can arise when collaborative tasks are set in conjunction. Proofs of the results are given in Appendix \ref{appendixa}.\par 
Intuitively, conflicting conjunctions of collaborative tasks in the fragment \eqref{eq:working fragment} occur when two or more collaborative tasks feature an intersection in the temporal domain (in terms of the intersection of their time intervals) which does not correspond to an intersection in the spatial domain (in terms of their truth sets).  We here propose a more general definition of conflicting conjunction compared to the one in \cite{marchesini2024communication}. We conjecture that these are the only five types of conflicting conjunctions over collaborative tasks according to fragment \eqref{eq:working fragment}. We acknowledge that conflicting conjunctions can also arise among conjunctions of independent and collaborative tasks, but we limit the presentation to conflicts over collaborative tasks. Adding conditions for conflicting conjunctions of independent and collaborative tasks is the subject of future work. \par
In the presentation of the different types of conflicting conjunctions, we consider the general conjunction of tasks $\phi_{ij}=\bigwedge_{k=1}^{K_{ij}}\varphi^{k}_{ij}$ over an edge $(i,j)$ as per \eqref{eq:conjunctions of multiple tasks}, with truth sets $\mathcal{B}_{ij}^k, \forall k =1, \ldots K_{ij}$ as per \eqref{eq:super level sets}. Moreover, let the set of indices $\mathcal{I}_G$, $\mathcal{I}_F$ such that $\mathcal{I}_G \cup \mathcal{I}_F = \{1,\ldots K_{ij}\}$ with
\begin{equation}\label{eq:always and eventually indices}
\begin{aligned}
\mathcal{I}_G = \{ k =1\ldots K_{ij} \;|\;\varphi^k_{ij} = G_{[a^k,b^k]} \mu_{ij}^k\}, \\
\mathcal{I}_F = \{ k =1\ldots K_{ij} \;|\;\varphi^k_{ij} = F_{[a^k,b^k]} \mu_{ij}^k\}.
\end{aligned}
\end{equation} 
While we present Facts 1-3 consecutively, Example \ref{second example} provides intuition on their meanings and implications.
\begin{Fact}(Conflict of type 1)\label{conflict 1}
Consider the conjunction of tasks $\phi_{ij}=\bigwedge_{k=1}^{K_{ij}}\varphi^{k}_{ij}$ and let $\mathfrak{L}_{ij}$ be a set of subsets of indices in $\mathcal{I}_G$ such that

\begin{equation}\label{eq:always sets intersections}
 \mathfrak{L}_{ij} := \{L \in 2^{\mathcal{I}_G}\setminus \{ \emptyset\}\;  |\;  \bigcap\nolimits_{l\in L} [a^{l},b^{l}] \neq \emptyset  \}. 
\end{equation}
If there exists a set of indices $L\in  \mathfrak{L}_{ij}$ such that $\bigcap_{l \in L} [a^{l},b^{l}] \neq \emptyset$ and  $\bigcap_{l \in L} \mathcal{B}_{ij}^{l}=\emptyset$ then $\phi_{ij}$ is a conflicting conjunction.
\end{Fact}
\par 
Concerning the intuitive meaning of conflicting conjunctions as per Fact \ref{conflict 1}, these occur any time a set of collaborative tasks, featuring an always operator, also feature a non-empty time interval intersection. Indeed, for any time in this time interval intersection, the relative state $\vec{e}_{ij}$ should be inside the truth set $\mathcal{B}_{ij}$ of all the considered tasks, which is possible only if the truth sets are intersecting as well. The next two facts involve conflicting conjunctions of tasks with mixed temporal operators.

\begin{Fact}(Conflict of type 2)\label{conflict 2}
Consider the conjunction of tasks $\phi_{ij}=\bigwedge_{k=1}^{K_{ij}}\varphi^{k}_{ij}$. For a given index $d\in \mathcal{I}_F$, let $\mathfrak{C}_{ij}(d)$ be the set of subsets of indices in $\mathcal{I}_G$ such that
\begin{equation}\label{eq:cover index set intersection}
\begin{aligned}
\mathfrak{C}_{ij}(d) = \{ C \in 2^{\mathcal{I}_G}\setminus \{ \emptyset\} \; | \: [a^{d},b^{d}] \subseteq \bigcup\nolimits_{l\in C} [a^{l},b^{l}] \;  \land \\ 
 \hspace{2cm}\; [a^{l},b^{l}] \cap [a^{d},b^{d}] \neq \emptyset, \forall l\in C   \}.
\end{aligned}
\end{equation}
Then, if  there exists $d\in \mathcal{I}_F$ and a set of indices $C \in \mathfrak{C}_{ij}(d)$ such that $\mathcal{B}_{ij}^{d}\cap\mathcal{B}_{ij}^{l}=\emptyset, \; \forall l\in C$ then $\phi_{ij}$ is a conflicting conjunction.
\end{Fact}

\begin{Fact}(Conflict of type 3)\label{conflict 3}
Consider the conjunction of tasks $\phi_{ij}=\bigwedge_{k=1}^{K_{ij}}\varphi^{k}_{ij}$. For a given index $d\in \mathcal{I}_F$, let $\mathfrak{D}_{ij}(d)$ be a set of subsets of indices in $\mathcal{I}_G$ such that
\begin{equation}\label{eq:cover index set union}
\mathfrak{D}_{ij}(d) = \{ D \in 2^{\mathcal{I}_G}\setminus \{ \emptyset\} \; | \: [a^{d},b^{d}] \subseteq \bigcap\nolimits_{l\in D} [a^{l},b^{l}] \}.
\end{equation}
Then, if there exists $d\in \mathcal{I}_F$ and a set of indices $D\in \mathfrak{D}_{ij}(d)$ such that $\mathcal{B}_{ij}^{d}\cap \bigcap_{l\in D}\mathcal{B}_{ij}^{l}=\emptyset$, then $\phi_{ij}$ is a conflicting conjunction.
\end{Fact}

\begin{figure}
     \centering
     \begin{subfigure}[b]{0.47\textwidth}
         \centering
         \includegraphics[width=\textwidth]{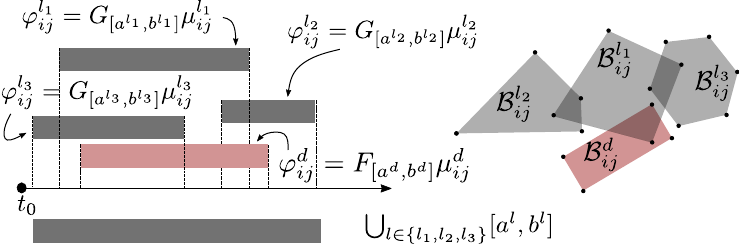}
         \caption{}
         \label{fig:upper conflict}
     \end{subfigure}
     \hfill
     \begin{subfigure}[b]{0.47\textwidth}
         \centering
         \includegraphics[width=\textwidth]{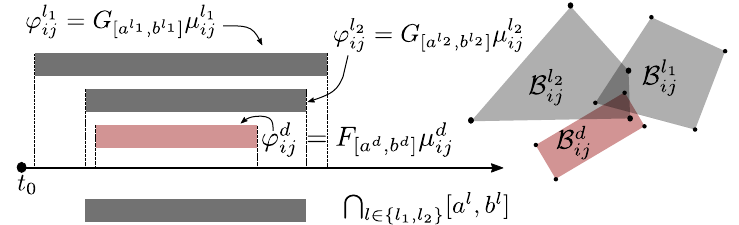}
         \caption{}
         \label{fig:lower conflict}
     \end{subfigure}
       \caption{Graphical exemplification of two conjunctions of tasks that do not induce conflicting tasks as per Fact \ref{conflict 2}. Detailed explanation in Example \ref{second example}.}
       \label{fig:graphical conflict}
\end{figure}
\par 
We provide some intuitions over the results of Facts \ref{conflict 2} and \ref{conflict 3}. Consider a collaborative task $\varphi_{ij}^d = F_{[a^d,b^d]}\mu_{ij}^d$, with $d\in \mathcal{I}_F$, such that its satisfaction can potentially occur at any time in the interval $[a^d,b^d]$ as per \eqref{eq:eventually old def}. Conflicts as per Fact \ref{conflict 2} can then occur if there exists a set of tasks $\phi_{ij}^l= G_{[a^l,b^l]}\mu_{ij}^l$ such that the union $\bigcup_{l}[a^l,b^l]$ is covering the interval $[a^d,b^d]$ in the sense that $\bigcup_{l}[a^l,b^l] \supseteq [a^d,b^d]$. In this case at least one of the tasks $\phi_{ij}^l$ must necessarily be satisfied together with $\varphi_{ij}^d$. On the other hand, conflicts as per Fact \ref{conflict 3} 
occur when the time interval intersection of the tasks $\varphi_{ij}^l$ is covering $[a^d,b^d]$ in the sense that $\bigcap_{l}[a^l,b^l] \supseteq [a^d,b^d]$. In this case, there exists at least one time instant $\tau\in [a^d,b^d]$ such that all tasks $\phi_{ij}^l= G_{[a^l,b^l]}\mu_{ij}^l$ must necessarily be satisfied together with $\varphi_{ij}^d = F_{[a^d,b^d]}\mu_{ij}^d$. The next example provides a graphical explanation for these intuitions.
\begin{example}\label{second example}
    Consider Figure \ref{fig:upper conflict} where a conjunction of four tasks $\phi_{ij} = \varphi_{ij}^{l_1} \land \varphi_{ij}^{l_2} \land \varphi_{ij}^{l_3} \land \varphi_{ij}^{d}$ is represented with $\varphi_{ij}^{l}=G_{[a^{l},b^{l}]}\mu_{ij}^{l},\; \forall l\in \{l_1,l_2,l_3\}$ and $\phi_{ij}^{d}=F_{[a^{d},b^{d}]}\mu_{ij}^{d}$. The left of Figure \ref{fig:upper conflict} shows the time interval intersection for the tasks, while the truth sets intersection is given on the right. This conjunction does not suffer from conflicting conjunctions as per Fact \ref{conflict 1}, \ref{conflict 2} or \ref{conflict 3}. Indeed, concerning conflicts of type \ref{conflict 1}, we have that $\mathfrak{L}_{ij} =\{ \{l_1,l_2\} ,\{l_1,l_3\},\{l_1\},\{l_2\},\{l_3\} \}$. First we note that that $[a^{l},b^{l}]\neq \emptyset$ and $\mathcal{B}_{ij}^{l}\neq \emptyset$  for all $l\in \{l_1,l_2,l_3\}$, thus avoiding conflicts as per type \ref{conflict 1} over the singleton sets in $\mathfrak{L}_{ij}$. Moreover, $[a^{l_1},b^{l_1}] \cap [a^{l_2},b^{l_2}] \neq \emptyset$ and $[a^{l_1},b^{l_1}] \cap [a^{l_3},b^{l_3}] \neq \emptyset$, while at the same time $\mathcal{B}_{ij}^{l_1}\cap \mathcal{B}_{ij}^{l_2} \neq \emptyset$ and  $\mathcal{B}_{ij}^{l_1}\cap \mathcal{B}_{ij}^{l_3} \neq \emptyset$ (Figure \ref{fig:upper conflict} on the right). Thus conflicts of type \ref{conflict 1} also do not arise for $\{l_1,l_2\},\{l_1,l_3\} \in \mathfrak{L}_{ij}$. Passing to conflicts of type \ref{conflict 2} we have  $\mathfrak{C}_{ij}(d) = \{\{l_1,l_2,l_3\}\}$ as per \eqref{eq:cover index set union}. From Figure \ref{fig:upper conflict} it is shown that $\mathcal{B}_{ij}^{l_1} \cap \mathcal{B}_{ij}^d \neq \emptyset$ thus avoiding the conflict of type \ref{conflict 2}. Finally turning to conflicts of type \ref{conflict 3}, we have that $\mathfrak{D}_{ij} = \emptyset $ such that no conflicts of this type case arise. Consider now Figure \ref{fig:lower conflict} where conjunction of three tasks $\phi_{ij} = \varphi_{ij}^{l_1} \land \varphi_{ij}^{l_2} \land \varphi_{ij}^{d}$ is represented in the same fashion as Figure \ref{fig:upper conflict}. Again $\{ l_1,l_2\} = \mathcal{I}_G$ while $\{d\} \in \mathcal{I}_d$. In this case, we focus on conflict of type \ref{conflict 3}, while the absence of conflicts of type \ref{conflict 1}, \ref{conflict 2} can be verified as in the previous example. Then we have $\mathfrak{D}_{ij}(d) = \{\{ l_1\},\{l_2\}, \{l_1,l_2\} \}$, as per \eqref{eq:cover index set intersection}, since the time intervals of the tasks $\varphi_{ij}^{l_1}$ and $\varphi_{ij}^{l_2}$, as well as their intersection, contain the time interval of the task $\varphi_{ij}^d$. Since the intersection $\mathcal{B}_{ij}^{l_1} \cap \mathcal{B}_{ij}^{l_2} \cap \mathcal{B}_{ij}^{l_d}$ is not empty (on the right in \ref{fig:lower conflict}), then conflicting conjunctions of type \ref{conflict 3} do not arise.
\end{example}
\par The next two conflicting conjunctions are defined over cycles of tasks in $\mathcal{G}_{\psi}$ rather than on a single edge. Namely, consider a cycle $\vec{\omega}$ over the task $\mathcal{G}_{\psi}$ and assume, for simplicity, that for each edge $(r,s) \in \epsilon(\vec{\omega})$ a single collaborative task $\phi_{rs} = \varphi_{rs}$ is defined. Then, the global task $\psi$ contains the conjunction of tasks $\land_{(r,s) \in \epsilon(\vec{\omega})} \varphi_{rs}$ and conflicting conjunction can arise if the cycle closure relation \eqref{eq:edge cycle} can not be satisfied together with the conjunction of tasks $\land_{(r,s) \in \epsilon(\vec{\omega})} \varphi_{rs}$. This intuition is formalised in the following two facts.
\begin{Fact}(Conflict of type 4)\label{conflict 4}
Consider a cycle $\vec{\omega}$ over $\mathcal{G}_{\psi}$ and the conjunction of tasks $\land_{(r,s) \in \epsilon(\vec{\omega})} \varphi_{rs}$ defined over $\vec{\omega}$ such that, with a slight abuse of notation, $\varphi_{rs}=G_{[a_{rs},b_{rs}]}\mu_{rs},\; \forall (r,s) \in \epsilon(\vec{\omega})$ with respective truth  sets $\mathcal{B}_{rs}, \forall (r,s)\in \epsilon(\vec{\omega})$ as per \eqref{eq:super level sets}. If $\bigcap_{(r,s)\in \epsilon(\vec{\omega})}[a_{rs},b_{rs}] \neq \emptyset$ and 
\begin{equation}\label{eq:Minkowski condition for conflict 3}
\vec{0} \not \in \bigoplus_{(r,s) \in \epsilon(\vec{\omega})}\mathcal{B}_{rs},
\end{equation}
then $\land_{(r,s) \in \epsilon(\vec{\omega})} \varphi_{rs}$ is a conflicting conjunction.
\end{Fact}
\begin{Fact}(Conflict of type 5)\label{conflict 5}
Consider a cycle $\vec{\omega}$ over $\mathcal{G}_{\psi}$ and the conjunction of tasks $\land_{(r,s) \in \epsilon(\vec{\omega})} \varphi_{rs}$ defined over $\vec{\omega}$ with respective truth sets $\mathcal{B}_{rs}, \forall (r,s)\in \epsilon(\vec{\omega})$ as per \eqref{eq:super level sets}. Furthermore, consider the single edge path $\vec{\pi}_i^p=[i,p]$ and the path $\vec{\pi}_p^i$ such that $\vec{\omega} = [\vec{\pi}_i^p, \vec{\pi}_p^i]$.  Let, without the loss of generality, $\varphi_{ip}=F_{[a_{ip},b_{ip}]}\mu_{ip}$ and $\varphi_{rs}=G_{[a_{rs},b_{rs}]}\mu_{rs}, \; \forall (r,s) \in \epsilon(\vec{\pi}_p^i)$. If it holds that $\cap_{(r,s)\in \epsilon(\vec{\pi}_p^i)}[a_{rs},b_{rs}] \supseteq [a_{ip},b_{ip}]$ and 
\begin{equation}\label{eq:Minkowski condition for conflict 4}
\vec{0} \not\in \mathcal{B}_{ip} \oplus \bigoplus_{(r,s) \in \epsilon(\vec{\pi}_i^p)}\mathcal{B}_{rs}
\end{equation}
then  $\land_{(r,s) \in \epsilon(\vec{\omega})} \varphi_{rs}$ is a conflicting conjunction.
\end{Fact}
\par 
While we provided the statement for conflicting conjunction of type \ref{conflict 4} and \ref{conflict 5}, we show in Section \ref{handling conflicting conjunctions} that these can not arise during our proposed decomposition thanks to the acyclicity assumption over the communication graph $\mathcal{G}_c$. We choose to report them as they represent an independent contribution to future work in this direction. The following assumption is considered further
\begin{assumption}\label{conflict free assumption}
    The original task graph $\mathcal{G}_{\psi}$ does not suffer from any conflicting conjunction as per Facts \ref{conflict 1}-\ref{conflict 5}.
\end{assumption}
This is a reasonable assumption over $\mathcal{G}_{\psi}$ since otherwise the task $\psi$ would be unsatisfiable even before the decomposition.
We explore in Section \ref{handling conflicting conjunctions} how the proposed decomposition approach preserves the conflict-free property passing from $\mathcal{G}_{\psi}$ to $\mathcal{G}_{\bar{\psi}}$.

\section{Task Decomposition}\label{task decomposition}
In this section, the decomposition approach to obtain tasks $\bar{\phi}^{ij}$ from the inconsistent tasks $\phi_{ij}$ defined over the edges $(i,j) \in \mathcal{E}_{\psi}\setminus \mathcal{E}_c$ is presented as per Problem \ref{problem1}. For the sake of clarity and to reduce the burden of notation, we present our results assuming that all the inconsistent tasks $\phi_{ij}$ do not contain conjunctions such that simply $\phi_{ij} = \varphi_{ij}$. Hence, by dropping the index $k$ in \eqref{eq:path specification} and replacing \eqref{eq:single decomposition} in \eqref{eq:conjunctions of decomposition}, we get that $\phi_{ij} = \varphi_{ij}$ is decomposed into the task $\bar{\phi}^{ij}= \land_{(r,s) \in \epsilon(\vec{\pi}_i^j)} \bar{\varphi}_{rs}^{ij}$. Note that this simplifying assumption is not required from our approach but is nonetheless introduced for the sake of clarity. Indeed, the same approach developed to decompose a single collaborative task  $\varphi_{ij}$ into an appropriate task $\bar{\phi}^{ij}$ as per \eqref{eq:single decomposition}, can be repeated for any index $k\in \{1,\ldots K_{ij}\}$ in the conjunction. With these considerations, in Sec. \ref{task decomposition subsection} the first main decomposition result is provided in the form of Lemma \ref{single formula decomposition}, where a 
procedure to define the task $\bar{\phi}^{ij}$ from an inconsistent task $\varphi_{ij}$ is derived. The task $\bar{\phi}^{ij}$ resulting from this decomposition is a function of a set of parameters that are later shown to be computable via convex optimization. Moreover,  Sec. \ref{handling conflicting conjunctions} and \ref{handling communication} detail how convex constraints over the former parameters can be enforced such that conflicting conjunctions are avoided and each task is communication consistent.
\subsection{Parametric tasks for decomposition}\label{task decomposition subsection}
Consider the communication inconsistent task $\phi_{ij} = \varphi_{ij}$ with $(i,j)\in\mathcal{E}_\psi\setminus\mathcal{E}_c$, where $h_{ij}= \langle -( A_{ij}(\vec{e}_{ij}-\vec{c}_{ij}) - \vec{z}_{ij}) \rangle$ is the predicate function associated with $\varphi_{ij}$ and $\mathcal{B}_{ij}=\mathcal{P}(\vec{A}_{ij},\vec{c}_{ij},\vec{z}_{ij})$ is the associated truth set as per \eqref{eq:predicates function standard form} and \eqref{eq:super level sets}. Furthermore, consider the following family of parametric predicate functions  
\begin{subequations}\label{eq:infty predicate function}
\begin{align}
\bar{h}^{ij}_{rs}(\vec{e}_{rs},\vec{\eta}_{rs}^{ij}) &:= \langle -(A_{ij}(\vec{e}_{rs}-\vec{c}^{ij}_{rs})-\alpha^{ij}_{rs}\vec{z}_{ij}) \rangle, \label{eq:parameteric predicate function} \\
\bar{\mathcal{B}}^{ij}_{rs}(\vec{\eta}_{rs}^{ij})&:= \{\vec{e}_{rs}\in \mathbb{X}_{rs}| \bar{h}^{ij}_{rs}(\vec{e}_{rs},\vec{\eta}_{rs}^{ij})\geq0\}, \label{eq:parameteric superlevel set}
\end{align}
\end{subequations}
such that
\begin{equation}\label{eq:parameteric predicate}
\bar{\mu}^{ij}_{rs} := \begin{cases}
\true& \; \text{if} \quad \bar{h}^{ij}_{rs}(\vec{e}_{rs},\vec{\eta}_{rs}^{ij})\geq 0 \\
\false& \; \text{if} \quad \bar{h}^{ij}_{rs}(\vec{e}_{rs},\vec{\eta}_{rs}^{ij})< 0,
\end{cases}
\end{equation}
where $\vec{\eta}_{rs}^{ij} =[\vec{c}_{rs}^{ij},\alpha_{rs}^{ij}]$ is a free parameter vector that needs to be optimised. The parameter vector $\vec{\eta}_{rs}^{ij}$ intuitively defines the position and the scale of the polytope $ \mathcal{P}(A_{ij},\vec{c}_{rs}^{ij},\alpha_{rs}^{ij}
\vec{z}_{ij}) = \bar{\mathcal{B}}^{ij}_{rs}(\vec{\eta}_{rs}^{ij})$ which is similar to $\mathcal{P}(A_{ij},\vec{c}_{ij},\vec{z}_{ij})$. 

\begin{figure}[!b]
    \centering
    \begin{subfigure}[b]{0.45\textwidth}
        \centering
        \includegraphics[width=\textwidth]{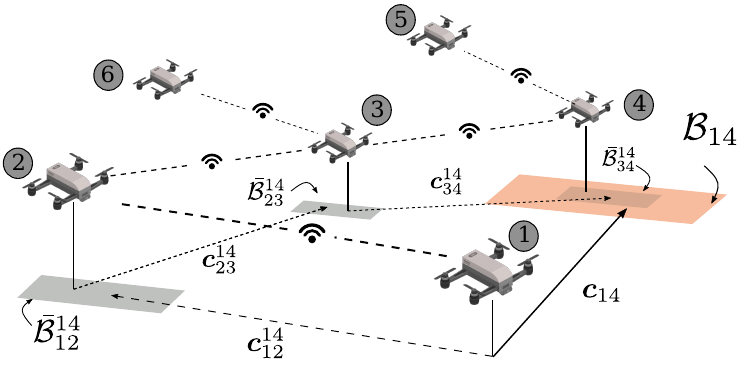}
        \caption{}
        \label{fig:drones}
    \end{subfigure}%
    \hfill
    \begin{subfigure}[b]{0.45\textwidth}
        \centering
        \includegraphics[width=\textwidth]{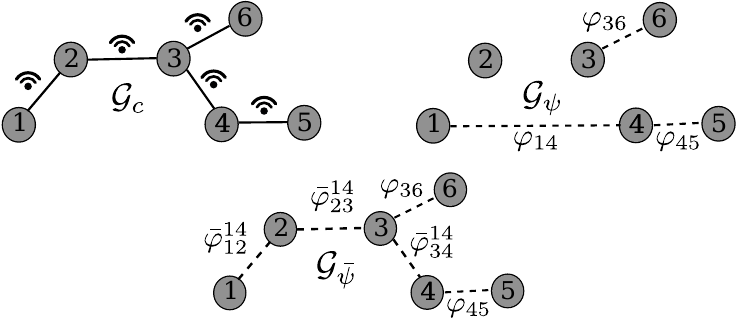}
        \caption{}
        \label{fig:graphs of the decompostion}
    \end{subfigure}
    \caption{Graphical representation of a task decomposition process according to Sec. \ref{task decomposition subsection} for a team of 6 drones. Detailed explanation is provided in Example \ref{drone main example}.}  
    \label{fig:drone example}
\end{figure}

From the family of parametric predicate functions in \eqref{eq:infty predicate function} a decomposition of the inconsistent task $\varphi_{ij}$ is obtained by a two-step procedure: 1) define a path of agents $\vec{\pi}_{i}^j$ from $i$ to $j$ through the communication graph $\mathcal{G}_c$; 2) define a set of \textit{parametric} tasks $\bar{\varphi}^{ij}_{rs} \; \forall (r,s) \in \epsilon(\vec{\pi}_i^j)$ with corresponding predicate functions $\bar{h}^{ij}_{rs}$ as per \eqref{eq:infty predicate function} such that $(\vec{x}(t),0)\models \bar{\phi}^{ij}= \bigwedge_{(r,s)\in \epsilon(\vec{\pi}_{i}^j)}\bar{\varphi}^{ij}_{rs} \Rightarrow (\vec{x}(t),0)\models\varphi_{ij}$. The solution to 1) is well studied in the literature so that we assume a path $\vec{\pi}_i^j$ between any pair $i,j\in \mathcal{V}$ can be found efficiently \cite[Ch. 2]{lavalle2006planning}. On the other hand, a solution to step 2) is provided by Lemma \ref{single formula decomposition}. 
\begin{lemma}\label{single formula decomposition}
Consider a collaborative task $\varphi_{ij}$ defined over the inconsistent edge $(i,j)\in \mathcal{E}_\psi \setminus \mathcal{E}_c$, with associated truth set $\mathcal{B}_{ij}=\mathcal{P}(A_{ij},\vec{c}_{ij},\vec{z}_{ij})$ as per \eqref{eq:super level sets}. Moreover, consider the path $\vec{\pi}_i^j$ through the communication graph $\mathcal{G}_c$ and $\bar{\phi}^{ij}=\bigwedge\nolimits_{(r,s)\in \epsilon(\vec{\pi}_{i}^j)}\bar{\varphi}_{rs}^{ij}$ as per \eqref{eq:path specification} such that 

\begin{equation}\label{eq:specific form of phi bar}
\bar{\varphi}^{ij}_{rs} = 
\begin{cases}
F_{[\bar{t},\bar{t}]}\bar{\mu}^{ij}_{rs} &\, \text{if} \quad \varphi_{ij}=F_{[a,b]}\mu_{ij}, \; \text{s.t} \; \bar{t}\in [a,b] \\
G_{[a,b]}\bar{\mu}^{ij}_{rs} &\, \text{if} \quad \varphi_{ij}=G_{[a,b]}\mu_{ij} ,
\end{cases}
\end{equation}
where $\bar{\mu}^{ij}_{rs}$, $\bar{h}^{ij}_{rs}(\vec{e}_{rs},\vec{\eta}_{rs}^{ij})$, $\mathcal{\bar{B}}^{ij}_{rs}(\vec{\eta}_{rs}^{ij})$ are as per \eqref{eq:infty predicate function}-\eqref{eq:parameteric predicate}.
If for each $(r,s)\in\epsilon(\vec{\pi}_i^j)$ the tasks $\bar{\varphi}^{ij}_{rs}$ are defined according to \eqref{eq:specific form of phi bar} and there exists parameters $\vec{\eta}_{rs}^{ij},\; \forall (r,s) \in \epsilon(\vec{\pi}_i^j)$ such that
 \begin{equation}\label{eq:minkosky inclusion}
 \bigoplus_{(r,s)\in\epsilon(\vec{\pi}_i^j)} \bar{\mathcal{B}}^{ij}_{rs}(\vec{\eta}_{rs}^{ij})\subseteq \mathcal{B}_{ij},
 \end{equation}
then $(\vec{x}(t),0)\models \bar{\phi}^{ij} \Rightarrow(\vec{x}(t),0)\models \varphi_{ij}$.
\end{lemma}
\begin{proof}
We prove the lemma for  $\varphi_{ij}:=F_{[a,b]}\mu_{ij}$ while the case of $\varphi_{ij}:=G_{[a,b]}\mu_{ij}$ follows a similar reasoning. We omit the dependency of $\bar{h}_{rs}^{ij},\bar{\mathcal{B}}_{rs}^{ij}$ with respect to $\vec{\eta}_{rs}^{ij}$ to reduce the notation. Given the path $\vec{\pi}_i^j$ over $\mathcal{G}_c$, then  $\bar{\phi}^{ij}$ is defined according to \eqref{eq:specific form of phi bar} as
$$
\bar{\phi}^{ij}=\bigwedge_{(r,s)\in \epsilon(\vec{\pi}_{i}^j)}\bar{\varphi}_{rs}^{ij}=\bigwedge_{(r,s)\in\epsilon(\vec{\pi}_i^j)}F_{[\bar{t},\bar{t}]}\bar{\mu}^{ij}_{rs},
$$
where $[\bar{t},\bar{t}]\subseteq[a,b]$. By \eqref{eq:conjunction robust} it holds $\rho^{\bar{\phi}^{ij}}(\vec{x}(t),0)=\min_{(r,s)\in\epsilon(\vec{\pi}_i^j)}\{\rho^{\bar{\varphi}_{rs}^{ij}}(\vec{x}(t),0)\}>0 \Leftrightarrow (\vec{x}(t),0)\models \bar{\phi}^{ij}$. Since, $\bar{\varphi}_{rs}^{ij}=F_{[\bar{t},\bar{t}]}\bar{\mu}^{ij}_{rs}$ for all $(r,s)\in \epsilon(\vec{\pi}_{i}^j)$, then by \eqref{eq:eventually robust} it holds, for each $(r,s)\in \epsilon(\vec{\pi}_i^j)$, that 
$\rho^{F_{[\bar{t},\bar{t}]} \mu_{rs}^{ij}} (\vec{x}(t),0)>0 \Leftrightarrow \min_{t\in [\bar{t},\bar{t}]}\bar{h}^{ij}_{rs}(\vec{e}_{rs}(t))> 0 \Leftrightarrow \bar{h}^{ij}_{rs}(\vec{e}_{rs}(\bar{t}))> 0$. Thus, by \eqref{eq:parameteric superlevel set}, $\vec{e}_{rs}(\bar{t})\underset{\text{\eqref{eq:parameteric superlevel set}}}{\in}\mathcal{\bar{B}}^{ij}_{rs},\; \forall (r,s)\in \epsilon(\vec{\pi}_i^j)$. On the other hand, from \eqref{eq:edge sequence} and \eqref{eq:minkosky inclusion} it holds at time $\bar{t}$ that 
$$
\vec{e}_{ij}(\bar{t})\underset{\text{\eqref{eq:edge sequence}}}{=}\sum_{(r,s)\in \epsilon(\vec{\pi}_i^j)}\vec{e}_{rs}(\bar{t})\in \bigoplus_{(r,s)\in \epsilon(\vec{\pi}_i^j)} \bar{\mathcal{B}}^{ij}_{rs} \underset{\text{\eqref{eq:minkosky inclusion}}}{\subseteq} \mathcal{B}_{ij}.
$$
Thus, for every satisfying trajectory $(\vec{x}(t),0)\models \bar{\phi}^{ij}$ it must hold $\vec{e}_{ij}(\bar{t})\in\mathcal{B}_{ij} \Rightarrow h_{ij}(\vec{e}_{ij}(\bar{t}))>0$. Since $\bar{t}\in [a,b]$, then by \eqref{eq:eventually robust} it holds $\rho^{\varphi^{ij}}(\vec{x}(t),0)=\min_{t\in[a,b]} (h_{ij}(\vec{e}_{ij}(t)))>0 \Rightarrow (\vec{x}(t),0)\models \varphi_{ij}$, which concludes the proof.
\end{proof}
 The following clarifying example is provided to build a graphical intuition of the result in Lemma \ref{single formula decomposition}.
\begin{example}\label{drone main example}
    Figure \ref{fig:drones} shows a system of 6 drones with states $\vec{x}_i=\vec{p}_i$ being the positions of each drone, while Figure \ref{fig:graphs of the decompostion} shows the task graph $\mathcal{G}_{\psi}$ and a communication graph $\mathcal{G}_c$ assigned to the agents. The edge $(1,4) \in \mathcal{E}_{\psi} \setminus \mathcal{E}_c$ is an inconsistent edge that requires decomposition through $\vec{\pi}_1^4 = [1,2,3,4]$ such that a new task $\bar{\phi}^{ij} = \bar{\varphi}_{12}^{14} \land \bar{\varphi}_{23}^{14} \land \bar{\varphi}_{34}^{14}$ is introduced as per Lemma \ref{single formula decomposition}. The new task graph resulting from the decomposition is then $\mathcal{G}_{\bar{\psi}}$ (bottom in Figure \ref{fig:graphs of the decompostion}). Figure \ref{fig:drones} also shows the similar truth sets $\bar{\mathcal{B}}_{rs}^{14},\; \forall (r,s)\in \epsilon(\vec{\pi}_1^4) = \{(1,2),(2,3),(3,4)\}$ (gray rectangles) and $\mathcal{B}_{14}$ (red rectangle). The inclusions relation \eqref{eq:minkosky inclusion} is graphically understood by letting the Minkowski sum $\bigoplus_{(r,s)\in \epsilon(\vec{\pi}_1^4)} \bar{\mathcal{B}}_{rs}^{14}$ to be a subset of $\mathcal{B}_{14}$.
\end{example}\par
\par
Note that when an inconsistent task $\varphi_{ij}$ featuring the eventually operator $F$ has to be decomposed, then the tasks $\bar{\varphi}_{rs}^{ij},\; \forall (r,s) \in \epsilon(\vec{\pi}_i^j)$ are necessarily synchronized over a common time instant $\bar{t}$ as specified by \eqref{eq:specific form of phi bar}. This synchronization is needed for the satisfaction of the original task $\varphi_{ij}$, although $\bar{t}$ can be chosen arbitrarily within $[a,b]$. Concerning the inclusions relation \eqref{eq:minkosky inclusion}, it is known from  Prop. \ref{sequence of polytopes} and Prop. \ref{polytopes inclusion} that \eqref{eq:minkosky inclusion} can be efficiently verified by a set of linear inequalities. Namely, consider again the truth set $\mathcal{B}_{ij}=\mathcal{P}(A_{ij},\vec{c}_{ij},\vec{z}_{ij})$ for the inconsistent task $\varphi_{ij}$ and let the decomposition path $\vec{\pi}_i^j$.  Furthermore, let the similar polytopes $\bar{\mathcal{B}}_{rs}^{ij}=P(A_{ij},\vec{c}_{rs}^{ij},\alpha_{rs}^{ij}\vec{z}_{ij}),\; \forall (r,s) \in \epsilon(\vec{\pi}_i^j)$ 
derived from the parametric task $\bar{\varphi}_{rs}^{ij}$ as per Lemma \ref{single formula decomposition} with parameters $\vec{\eta}_{rs}^{ij} = [\vec{c}_{rs}^{ij},\alpha_{rs}^{ij}]$. If we let the generator set $V_{ij} =\nu(\mathcal{P}(A_{ij},\vec{0},\vec{z}_{ij}))$, then the inclusion relation \eqref{eq:minkosky inclusion} is verified if it holds
\begin{equation}\label{eq: linear inclusion}
\sum_{(r,s)\in \epsilon(\vec{\pi}_i^j)} \Biggl( M_{ij}\vec{\eta}^{ij}_{rs} - \frac{Z_{ij}}{|\epsilon(\vec{\pi}_i^j)|}\begin{bmatrix}\vec{c}_{ij}\\1\end{bmatrix}\Biggr) \leq \vec{0},
\end{equation}
where
\begin{equation}\label{eq:definition of the intersection inclusion matrix}
M_{ij} = \begin{bmatrix}A_{ij} G_1\\ \vdots \\ A_{ij} G_{|V_{ij}|}\end{bmatrix}, \quad  Z_{ij} = \vec{1}_{|V_{ij}|} \otimes \comp{A_{ij}}{\vec{z}_{ij}},
\end{equation}
and $G_k = \comp{I_n}{\vec{v}_k} \in \mathbb{R}^{n\times (n+1)}$, $\vec{v}_k \in V_{ij},\, \forall k = 1,\ldots |V_{ij}|$. At the same time, it is known from Prop. \ref{inclusion accuracy} that if \eqref{eq:minkosky inclusion} holds, then the satisfaction of \eqref{eq:minkosky inclusion} implies
\begin{equation}\label{eq:task decomposition accuracy}
    \sum_{(r,s) \in \epsilon(\vec{\pi}_i^j)} \alpha_{rs}^{ij} \leq 1,
\end{equation}
where \eqref{eq:task decomposition accuracy} holds with equality if \eqref{eq:minkosky inclusion} does. We refer to the left-hand side of \eqref{eq:task decomposition accuracy} as the decomposition accuracy of the decomposed task $\varphi_{ij}$. We show in Section \ref{optimization section} that maximising the decomposition accuracy on the right-hand side of \eqref{eq:task decomposition accuracy} is indeed the main objective of our decomposition approach. Next, in Section \ref{handling communication} and \ref{handling conflicting conjunctions}, we provide a set of constraints that are imposed over the parameters $\vec{\eta}_{rs}^{ij}$ such that the respective tasks $\bar{\varphi}_{rs}^{ij}$ are communication consistent and such that the tasks $\bar{\varphi}_{rs}^{ij}$ do not cause conflicts as per Facts \ref{conflict 1}-\ref{conflict 5}.

\subsection{Handling conflicting conjunctions during decomposition}\label{handling conflicting conjunctions}
\begin{figure}
    \centering
    \includegraphics[width=.45\textwidth]{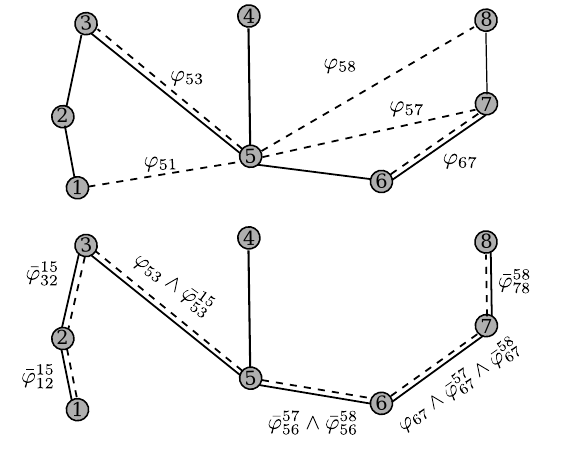}
    \caption{Graphical decomposition of formulas $\varphi_{58}$, $\varphi_{57}$, $\varphi_{51}$. (Top) Initial task graph (dashed lines) and communication graph (solid). (Bottom) Final task graph (dashed lines) and communication graph (solid).}
    \label{fig:illustrative decomposition}
\end{figure}

In this section, it is clarified how conflicting conjunctions as per Facts \ref{conflict 1}-\ref{conflict 5} might arise during decomposition and how to resolve them by appropriately constraining the parameters $\vec{\eta}_{rs}^{ij}$ associated with each $\bar{\varphi}_{rs}^{ij}$. Namely, we analyze how the consistent and conflict-free collaborative tasks $\phi_{rs}$ for all $(r,s)\in \mathcal{E}_{\pi} \subseteq \mathcal{E}_c$, that were in place \textit{before} the decomposition, are modified by the introduction of the new tasks $\bar{\varphi}_{rs}^{ij},\; \forall (i,j) \in \mathcal{E}_{\psi} \setminus \mathcal{E}_c$ resulting from the decomposition, and how the consistency and conflict-free properties for these can be preserved. As in the previous sections, it is assumed that inconsistent tasks are without conjunctions as $\phi_{ij}=\varphi_{ij}$. A change of notation is required at this point of the presentation and applies hereafter. Namely, let the sets
\begin{subequations}
\begin{align}
\mathcal{E}_{\pi} &:= \bigcup_{(i,j)\in \mathcal{E}_{\psi}\setminus \mathcal{E}_c} \epsilon(\vec{\pi}_i^j)\label{eq:definition of theta},\\
\Pi_{rs}&:= \{(i,j) \in \mathcal{E}_{\psi} \setminus \mathcal{E}_{c} \; |\; (r,s) \in \epsilon(\vec{\pi}_i^j)\}\label{eq:definition of pi edge},
\end{align}
\end{subequations}
where the set $\mathcal{E}_{\pi} \subseteq \mathcal{E}_{c}$ contains all the edges in $\mathcal{G}_c$ that are used for the decomposition of all the inconsistent tasks $\phi_{ij}$ with $(i,j)\in \mathcal{E}_{\psi}\setminus \mathcal{E}_c$, while $\Pi_{rs}\subseteq \mathcal{E}_{\psi}\setminus \mathcal{E}_c$ contains all the inconsistent edges $(i,j) \in \mathcal{E}_{\psi} \setminus \mathcal{E}_{c}$ such that the decomposition path $\vec{\pi}_i^j$ is passing though $(r,s)\in \mathcal{E}_{\pi} \subseteq \mathcal{E}_c$. Moreover, let the collaborative task $\phi_{rs}:= \land_{k=1}^{K_{rs}} \varphi_{rs}^{k}$, as per \eqref{eq:conjunctions of multiple tasks}, be defined over $(r,s)\in \mathcal{E}_{\pi}$ \textit{\textbf{before}} the decomposition. Then, \textit{\textbf{after}} the decomposition of all the inconsistent tasks $\varphi_{ij},\; \forall (i,j) \in \mathcal{E}_{\psi} \setminus \mathcal{E}_{c}$, we have that an additional conjunction of parametric tasks given by $\land_{(i,j) \in \Pi_{rs}}\bar{\varphi}_{rs}^{ij}$ is included over the edge $(r,s) \in \mathcal{E}_{\pi}$. Let then
\begin{equation}\label{eq:parametric set of indices}
\bar{\mathcal{K}}_{rs} = \{K_{rs}+1,\ldots K_{rs} +  |\Pi_{rs}|\},
\end{equation}
and let the one-to-one index mapping
\begin{equation}\label{eq:edge mapping for new tasks}
y_{\Pi_{rs}} : \Pi_{rs} \rightarrow \bar{\mathcal{K}}_{rs},
\end{equation}
that maps each unique $(i,j) \in \Pi_{rs}$ (an inconsistent edge whose decomposition path $\vec{\pi}_i^j$ passes though $(r,s)$) to a unique index in $\bar{\mathcal{K}}_{rs}$ so that the following notation equivalence is established   
\begin{equation}
    \bigwedge_{(i,j) \in \Pi_{rs}} \bar{\varphi}^{ij}_{rs} =  \bigwedge_{k\in \bar{\mathcal{K}}_{rs}} \varphi^{k}_{rs} ,
\end{equation}
where we dropped the bar notation over the parametric tasks to ease the notation. Hence,  \textbf{\textit{after}} decomposition, the tasks $\phi_{rs}$ for each $(r,s)\in \mathcal{E}_{\pi}$ take the form
\begin{equation}\label{eq:overloading}
\phi_{rs} := \underbrace{\bigwedge_{k=1}^{K_{rs}}\varphi_{rs}^k}_{\text{before decompositon}} \land \underbrace{\bigwedge_{k \in \bar{\mathcal{K}}_{rs}}\varphi^{k}_{rs}}_{\text{parametric tasks}} = \bigwedge_{k\in \mathcal{K}_{rs}}\varphi^{k}_{rs} ,
\end{equation}
where $\mathcal{K}_{rs} = \{1, \ldots K_{rs} \} \cup \bar{\mathcal{K}}_{rs}$. A clarifying example is provided.
\begin{example}\label{example indexing}
Consider Figure \ref{fig:illustrative decomposition}. The upper panel represents the communication graph (solid edges) together with the task graph (dashed edges) such that tasks $\varphi_{58},\; \varphi_{57}$ and $\varphi_{51}$ are inconsistent. The consistent edge $(6,7)\in \mathcal{E}_c$ is on the decomposition paths $\vec{\pi}_5^8=[5,6,7,8]$ and  $\vec{\pi}_5^7=[5,6,7]$ for both inconsistent tasks $\varphi_{58}$ and $\varphi_{57}$, while the task $\phi_{67} = \varphi_{67}$ is assigned to the edge \textit{before} the decomposition. From the index mapping $y_{\Pi_{67}}$, with  $y_{\Pi_{67}}((5,7)) = 2$ and  $y_{\Pi_{67}}((5,8)) = 3$ it derives
$$
\varphi_{67} \land \varphi_{67}^{57} \land \varphi_{67}^{58} \underset{\text{$y_{\Pi_{rs}}(\cdot)$}}{=} \varphi_{67}^1   \land  \varphi_{67}^2  \land \varphi_{67}^3,
$$
 where the index $1$ was assigned to the first task for consistency with the fact that $K_{rs}=1$. The final task graph and communication graphs are then represented by the lower panel in Figure \ref{fig:illustrative decomposition}.
 \end{example}\par 
 With this new notation, this section is concerned with stating sufficient conditions on the parameters $\vec{\eta}_{rs}^{k}, \; \forall k \in \bar{\mathcal{K}}_{rs}$ associated with the parametric tasks $\varphi_{rs}^k, \; \forall k \in \bar{\mathcal{K}}_{rs}$ such that the task $\phi_{rs}$ in \eqref{eq:overloading} is not a conflicting conjunction. Indeed, we highlight again that after the introduction of the index map $y_{\Pi_{rs}}$, the tasks $\varphi_{rs}^k, \; \forall k \in \bar{\mathcal{K}}_{rs}$ are as per \eqref{eq:specific form of phi bar}, with corresponding truth set given by $\mathcal{B}_{rs}^k=\mathcal{P}(A_{rs}^k,\vec{c}_{rs}^k,\alpha_{rs}^k\vec{z}^k_{rs})$ as per \eqref{eq:parameteric superlevel set} and such that $\vec{\eta}_{rs}^k:= [\vec{c}_{rs}^{k}, \alpha_{rs}^k]$ is a parameter to be optimised. On the other hand, the tasks $\varphi_{rs}^k,\; k\in \{1,\ldots K_{rs}\}$ are \textit{not} parametric, as per \eqref{eq:multi agent spec temporal}, and such that their truth set is a fixed polytope in the same general form $\mathcal{B}_{rs}^k=\mathcal{P}(A_{rs}^k,\vec{c}_{rs}^k,\alpha_{rs}^k\vec{z}^k_{rs})$, but with $\alpha_{rs}^k=1$ and $\vec{\eta}_{rs}^k=[\vec{c}_{rs}^{k},1]$ having fixed value as per \eqref{eq:super level sets}. Note that, in general, the polytopes $\mathcal{B}_{rs}^k, \forall k\in \mathcal{K}_{rs}$ are \textit{not} similar (cf. Def \ref{similar}). We are now ready to present the next main results. Following the presentation of the different types of conflicting conjunctions, we often refer to the set of indices $\mathcal{I}_G$ and $\mathcal{I}_F$ as per \eqref{eq:always and eventually indices} where  $\mathcal{I}_G \cup \mathcal{I}_F = \mathcal{K}_{rs}$,  $\varphi^k_{rs} = G_{[a^k,b^k]} \mu_{rs}^k, \; \forall k\in \mathcal{I}_G$ and $\varphi^k_{rs} = F_{[a^k,b^k]} \mu_{rs}^k, \; \forall k\in \mathcal{I}_F$.
\begin{proposition}\label{conflict resolution 1}
Consider $\phi_{rs} = \bigwedge_{k\in\mathcal{K}_{rs}} \varphi^{k}_{rs}$ as per \eqref{eq:overloading} and the set of subset of indices $\mathfrak{L}_{rs}$ as per \eqref{eq:always sets intersections} over the edge $(r,s)$. Furthermore, let $\mathfrak{L}^{M}_{rs} \subseteq \mathfrak{L}_{rs}$ be defined as $\mathfrak{L}^{M}_{rs} =\{ L \in \mathfrak{L}_{rs} \; |  \; L \not \subset L',  \, \forall L' \in \mathfrak{L}\}$. Then if for each set of indices $L \in \mathfrak{L}^{M}_{rs}$ there exists $\vec{\xi}\in \mathbb{R}^n$, such that
\begin{equation} \label{eq:proposition intersection 1}
\begin{aligned}
    A_{rs}^{l}\vec{\xi} -\comp{A_{rs}^{l}}{\vec{z}_{rs}^{l}} \vec{\eta}^{l}_{rs} \leq \vec{0}, \forall l\in L\\
    \end{aligned}
\end{equation}
then  $\phi_{rs}$ is not a conflicting conjunction as per Fact \ref{conflict 1}.
\end{proposition}
\begin{proof}
By Prop. \ref{polytopes inclusion} we know that the set of inequalities \eqref{eq:proposition intersection 1} guarantees that for every $L\in \mathfrak{L}^{M}_{rs}$ it holds $\bigcap_{l\in L}\mathcal{B}_{rs}^{l}\supseteq \{\vec{\xi}\} \neq \emptyset$, thus disproving the conditions for conflicting conjunction over the sets of indices $L_{rs}\in \mathfrak{L}^M_{rs}$ as per Fact \ref{conflict 1}. Consider now a set $L'$ such that $L'\in \mathfrak{L}_{rs} \setminus \mathfrak{L}^M_{rs}$. Then necessarily, by definition of $\mathfrak{L}^M_{rs}$, there must be an $L\in\mathfrak{L}^M_{rs}$ such that $L'\subset L$. We can then write $\bigcap_{l\in L'}\mathcal{B}_{rs}^{l} \supseteq \bigcap_{l\in L}\mathcal{B}_{rs}^{l}\supseteq \{\vec{\xi}\} \neq \emptyset$. Since $L'$ was chosen arbitrarily from  $\mathfrak{L}_{rs}$, then the conditions for conflicting conjunction as per Facts \ref{conflict 1} do not hold for any $L'\in \mathfrak{L}_{rs}$ concluding the proof.
\end{proof} 
\par The set $\mathfrak{L}^{M}_{rs}$ represents the set of subsets in  $\mathfrak{L}_{rs}$ that are ``maximal", in the sense that any set of indices $L\in \mathfrak{L}^{M}_{rs}$ is not contained in any other set $L' \in \mathfrak{L}_{rs}$. Therefore the result of Prop. \ref{conflict resolution 1} suggests that checking for conflicting conjunctions over the subset of indices in $\mathfrak{L}^{M}_{rs}$ is sufficient to avoid conflicting conjunctions of type \ref{conflict 1}, instead of checking all the possible subset of indices in $\mathfrak{L}_{rs}$ where $|\mathfrak{L}^{M}_{rs}| \leq |\mathfrak{L}_{rs}|$. Second, once the power set of indices $2^{\mathcal{I}_G}$ is computed, then extracting the set $\mathfrak{L}_{rs}\subseteq 2^{\mathcal{I}_G}$ (and thus $\mathfrak{L}^{M}_{rs}$) is computationally efficient as it requires only to verify the condition $\max_{l\in L} a^l \leq \min_{l\in L} b^l$ for each set $L\in 2^{\mathcal{I}_G}$. Third, only the sets of indices $L\in \mathfrak{L}^{M}_{rs}$ such that $L \cap \bar{\mathcal{K}}_{rs} \neq \emptyset$ have to be checked for conflicting conjunctions since, by Assumption \ref{conflict free assumption}, the conjunction $\land_{l\in \{1, \ldots K_{rs} \}} \varphi_{rs}^l$ as per \eqref{eq:overloading} is not a conflict of type \ref{conflict 1}. 
\begin{proposition}\label{conflict resolution 2}
Consider $\phi_{rs} = \bigwedge_{k\in\mathcal{K}_{rs}} \varphi^{k}_{rs}$ as per \eqref{eq:overloading} and the set of subset of indices $\mathfrak{C}_{rs}(d)$ as per \eqref{eq:cover index set union}. Furthermore let  $\mathfrak{C}_{rs}^{m}(d) \subseteq \mathfrak{C}_{rs}(d)$ be defined as $\mathfrak{C}_{rs}^{m}(d) =\{ C \in \mathfrak{C}_{rs}(d) \; |  \; C \not \supset C',  \, \forall C' \in \mathfrak{C}_{rs}(d)\}$. If for every $d \in \mathcal{I}_F$ and for every $C\in \mathfrak{C}_{rs}^{m}(d)$ there exists $l\in C$ such that 
\begin{equation} \label{eq:proposition intersection 2a}
\begin{aligned}
    A_{rs}^{l}\vec{\xi} -\comp{A_{rs}^{l}}{\vec{z}_{rs}^{l}} \vec{\eta}^{l}_{rs} \leq \vec{0},  \\
    A_{rs}^{d}\vec{\xi} -\comp{A_{rs}^{d}}{\vec{z}_{rs}^{d}} \vec{\eta}^{d}_{rs} \leq \vec{0}, \\
    \end{aligned}
\end{equation}
for some vector $\vec{\xi} \in \mathbb{R}^n$, then  $\phi_{rs}$ is not a conflicting conjunction as per Fact \ref{conflict 2}.
\end{proposition}
\begin{proof}
For a given $d\in \mathcal{I}_F$ consider the set $C\in \mathfrak{C}_{rs}^{m}(d) \subset \mathfrak{C}_{rs}(d)$ such that, by definition of $\mathfrak{C}_{rs}(d)$ in \eqref{eq:cover index set union}, we have $[a^{d},b^{d}] \subseteq \bigcup_{l\in C} [a^{l},b^{l}]$. Then by equation Prop. \ref{polytopes inclusion} it is known that \eqref{eq:proposition intersection 2a} ensures the existence of at least one index $l\in C$ such that $\mathcal{B}_{rs}^{d} \cap \mathcal{B}_{rs}^{l} \supseteq \{\vec{\xi}\} \neq \emptyset$. Therefore, a conflict as per Fact \ref{conflict 2} is avoided for the index set C. Now consider another index set $C' \neq C$ such that $C'\in \mathfrak{C}_{rs}(d) \setminus \mathfrak{C}_{rs}^{m}(d)$. By definition of $\mathfrak{C}_{rs}^{m}(d)$, it also holds $C \subset C'$. It is now sufficient to note that the satisfaction \eqref{eq:proposition intersection 2a} for one index $l\in C \subset C'$  disproves again the condition for conflicting conjunction as per Fact \ref{conflict 2} over $C'$. Since the result is valid for any $d\in \mathcal{I}_F$, any $C \in \mathfrak{C}_{rs}^m(d)$ and any $C' \in \mathfrak{C}_{rs}(d)\setminus \mathfrak{C}_{rs}^m(d)$, we have proved that conflicts as per type \ref{conflict 2} do not arise over $\phi_{rs}$.
\end{proof}
\par 
Similarly to Prop. \ref{conflict resolution 1}, the set $\mathfrak{C}^{m}_{rs}(d)$ represents the set of subsets in  $\mathfrak{C}_{rs}(d)$ for a given $d\in \mathcal{I}_{F}$, that is ``minimal", in the sense that any set of indices $C\in \mathfrak{C}^{m}_{rs}$ does not contain any set $C' \in \mathfrak{C}_{rs}(d)$ thus reducing the number of subsets of indices of tasks to be checked for conflicting conjunctions. The same remarks highlighted for Prop. \ref{conflict resolution 1} are also valid for Prop. \ref{conflict resolution 2}. The last result of this section is presented next.
\begin{proposition}\label{conflict resolution 3}
Consider $\phi_{rs} = \bigwedge_{k\in\mathcal{K}_{rs}} \varphi^{k}_{rs}$ as per \eqref{eq:overloading}  and the set of subset of indices $\mathfrak{D}_{rs}(d)$ as per \eqref{eq:cover index set intersection}. Furthermore, let $\mathfrak{D}_{rs}^{M}(d) \subseteq \mathfrak{D}_{rs}(d)$ be defined as $\mathfrak{D}_{rs}^{M}(d) =\{ D \in \mathfrak{D}_{rs}(d) \; |  \; D \not \subset D',  \, \forall D' \in \mathfrak{D}_{rs}(d)\}$. If for every $d \in \mathcal{I}_F$ and for every $D\in \mathfrak{D}_{rs}^{M}(d)$ it holds 
\begin{equation} \label{eq:proposition intersection 2b}
\begin{aligned}
    &A_{rs}^{l}\vec{\xi} -\comp{A_{rs}^{l}}{\vec{z}_{rs}^{l}} \vec{\eta}^{l}_{rs} \leq \vec{0}, \; \forall l\in D ,   \\
    &A_{rs}^{d}\vec{\xi} -\comp{A_{rs}^{d}}{\vec{z}_{rs}^{d}} \vec{\eta}^{d}_{rs} \leq \vec{0}, \\
    \end{aligned}
\end{equation}
for some vector $\vec{\xi} \in \mathbb{R}^n$, then  $\phi_{rs}$ is not a conflicting conjunction as per Fact \ref{conflict 3}.
\end{proposition}
\begin{proof}
For a given $d\in \mathcal{I}_F$ consider the set $D\in \mathfrak{D}_{rs}^{M}(d) \subset \mathfrak{D}_{rs}(d)$ such that, by definition of $\mathfrak{D}_{rs}(d)$ in \eqref{eq:cover index set intersection}, we have $[a^{d},b^{d}] \subseteq \bigcap_{l\in D} [a^{l},b^{l}]$. Then by equation Prop. \ref{polytopes inclusion} it is known that \eqref{eq:proposition intersection 2b} ensures that $\mathcal{B}_{rs}^{d} \cap \bigcap_{l\in D} \mathcal{B}_{rs}^{l} \supseteq \{\vec{\xi}\}$. Therefore, a conflict as per Fact \ref{conflict 3} is avoided over $D$. Now consider another set of indices $D' \neq D$ such that $D'\in \mathfrak{D}_{rs}(d) \setminus \mathfrak{D}_{rs}^{M}(d)$ and, by definition, it holds $D \supset D'$. Then the condition \eqref{eq:proposition intersection 2b} directly disproves the condition for conflicting conjunction as per Fact \ref{conflict 3} since, it holds $\mathcal{B}_{rs}^{d} \cap \bigcap_{l\in D'}\mathcal{B}_{rs}^{l} \supseteq \mathcal{B}_{rs}^{d} \cap \bigcap_{l\in D}\mathcal{B}_{rs}^{l} \supseteq \{\vec{\xi}\}$. Since the argument is valid for any $d\in \mathcal{I}_{F}$, any $D\in \mathfrak{D}_{rs}(d)$ and any $D'\in \mathfrak{D}_{rs}(d) \setminus \mathfrak{D}_{rs}^{M}(d)$, we have proved that conflicts as per type \ref{conflict 3} do not arise over $\phi_{rs}$.
\end{proof}\par
One last definition provides a compact and uniform way to verify the absence of any type of conflicting conjunctions over $\phi_{rs} = \land_{k\in \mathcal{K}_{rs}} \varphi_{rs}^k$ as per Prop. \ref{conflict resolution 1}, \ref{conflict resolution 2} and \ref{conflict resolution 3}. Namely, let
\begin{equation}\label{eq:the set of indices for conflicts}
\begin{aligned}
\mathfrak{Q}_{rs} &= \bigl( \bigcup_{d\in \mathcal{I}_F} \{ \; \{d\} \cup \{l\} \;|\; l \in C, \;  \forall C\in \mathfrak{C}_{rs}^m(d)\} \bigr) \cup \\
&\hspace{0.4cm}\bigl( \bigcup_{d\in \mathcal{I}_F} \{\; \{d\} \cup D \;|\; \forall D\in \mathfrak{D}_{rs}^M(d)\} \bigr) \cup \mathfrak{L}^{M}_{rs},
\end{aligned}
\end{equation}
and let  
\begin{subequations}
\begin{equation}\label{eq:auxiliary set}
\Xi_{rs} = \{ \vec{\xi}_{rs}^q ,\, \forall q = 1,\ldots |\mathfrak{Q}_{rs}|\}\\
\end{equation}
\begin{equation}\label{eq:set q index mapping}
y_{\mathfrak{Q}_{rs}} : \mathfrak{Q}_{rs} \mapsto \{1, \ldots |\mathfrak{Q}_{rs}| \},
\end{equation}
\end{subequations}
 where $\Xi_{rs}$ is the set of auxiliary variables applied to impose the conditions \eqref{eq:proposition intersection 1},  \eqref{eq:proposition intersection 2a} and \eqref{eq:proposition intersection 2b}, while $y_{\mathfrak{Q}_{rs}}$ is the index mapping relating each set of indices $Q\in \mathfrak{Q}$ with an auxiliary variable $\vec{\xi}_{rs}^q$ such that $q= y_{\mathfrak{Q}_{rs}}(Q)$. With these new definitions, the conjunctions $\wedge_{l\in Q}\varphi_{rs}^l$ for all $Q\in \mathfrak{Q}_{rs}$ represent potential conflicting conjunctions over the edge $(r,s) \in \mathcal{E}_{\pi}$. To avoid conflicts, the satisfaction of \eqref{eq:proposition intersection 1}, \eqref{eq:proposition intersection 2a} and \eqref{eq:proposition intersection 2b} over $(r,s)\in \mathcal{E}_{\pi}$ can be compactly written as 
\begin{equation}\label{eq:elegant conflict resolution equation}
    A_{rs}^l \vec{\xi}_{rs}^{q} - \comp{A_{rs}^{l}}{\vec{z}_{rs}^{l}} \vec{\eta}^{l}_{rs} \leq \vec{0}, \;  \forall Q\in \mathfrak{Q}_{rs}, \;  \forall l\in Q,  
\end{equation}
such that $q= y_{\mathfrak{Q}_{rs}}(Q)$.\par
It now remains to clarify how conflicting conjunctions as per Fact \ref{conflict 4}-\ref{conflict 5} can be avoided. In this respect, Assumption. \ref{acyclic communication} ensures that cycles of tasks as per Fact \ref{conflict 4}-\ref{conflict 5} can not arise during decomposition of $\mathcal{G}_{\psi}$ into $\mathcal{G}_{\bar{\psi}}$. Indeed, since only edges in the communication graph $\mathcal{G}_c$ are applied for the decomposition and since $\mathcal{G}_{c}$ is acyclic, then, by construction, the new task graph $\mathcal{G}_{\bar{\psi}} \subseteq \mathcal{G}_{c}$ must necessarily be acyclic, for which conflicts of types \ref{conflict 4} and \ref{conflict 5} do not arise. The reason why conditions \eqref{eq:Minkowski condition for conflict 3}-\eqref{eq:Minkowski condition for conflict 4} are excluded in our decomposition approach is because their enforcement leads to non-convex constraints over the parameters $\vec{\eta}$ as briefly explained next. Consider, for consistency with the presentation of Fact \ref{conflict 4}-\ref{conflict 5}, a conjunction of tasks $\land_{(r,s) \in \vec{\omega}}\varphi_{rs}$ for some cycle $\vec{\omega}$ and let the associated  non-similar polytopic truth sets $\mathcal{B}_{rs} = \mathcal{P}(A_{rs}, \vec{c}_{rs}, \alpha_{rs},\vec{z}_{rs}),\; \forall (r,s) \in \epsilon(\vec{\omega})$ with $\vec{\eta}_{rs}= [\vec{c}_{rs}, \alpha_{rs}]$ being parameters to be optimised. Then, let for brevity, $V_{rs} = \nu(\mathcal{B}_{rs}) =\{\vec{v}_{rs}^k\}_{k=1}^{|V_{rs}|},\; \forall (r,s)\in \epsilon(\vec{\omega})$ where, by Prop. \ref{linear scaling and translation} we have that each vector $\vec{v}^k_{rs}(\vec{\eta}_{rs})\in V_{rs}$ is expressed as a linear function in $\vec{\eta}_{rs}$. Moreover, let $V^{\oplus} = \bigoplus_{(r,s) \in \epsilon(\vec{\omega})} V_{rs} = \{\vec{v}_{\oplus}^k(\vec{\eta}_{\oplus})\}_{k=1}^{|V^{\oplus}|}$, such that $\vec{\eta}_{\oplus} = [\vec{\eta}_{rs}]_{(r,s) \in \epsilon(\vec{\omega})}$, and such that each vector $\vec{v}_{\oplus}^k(\vec{\eta_{\oplus}}) \in V^{\oplus}$ depends linearly on $\vec{\eta}_{\oplus}$. Noting that $\bigoplus_{(r,s) \in  \epsilon(\vec{\omega})}\mathcal{B}_{rs} = conv(\bigoplus_{(r,s) \in \epsilon(\vec{\omega})} V_{rs})$ \cite[Ch. 15, pp. 316]{gruber1993handbook}, the definition of convex hull in Def. \ref{convex hull} can be applied to enforce the inclusion $\vec{0}\in \bigoplus_{(r,s) \in \epsilon(\vec{\omega})} \mathcal{B}_{rs}$ by introducing the coefficients $\lambda_{k} \geq 0,\; k =1, \ldots |V^{\oplus}|$ with  $\sum_{k=1}^{|V^{\oplus}|}\lambda_{k}=1$ and such that $\sum_{k=1}^{|V^{\oplus}|}\lambda_{k} \vec{v}^k_{\oplus}(\vec{\eta}_{\oplus}) = \vec{0}$. However, this last summation is a non-convex constraint in the variables $\lambda_i, \forall k=1, \ldots |V^{\oplus}|$ and $\vec{\eta}_{\oplus}$. Moreover, the decentralization property of our decomposition approach, as presented in Section \ref{distributed task decompostion section}, is lost if this last type of constraint is introduced. When the acyclicity assumption on $\mathcal{G}_c$ can not be enforced, a solution to avoid conflicting conjunctions of type 4 and 5 was provided in our previous work \cite{marchesini2024communication} by restricting the sets $\mathcal{B}_{rs}$ to be hyper-rectangles instead of general polytopes. \par
\subsection{Handling communication consistency}\label{handling communication}
Concerning communication consistency of the parametric tasks $\varphi_{rs}^l,\; \forall l\in\bar{\mathcal{K}}_{rs}$ over the edges $(r,s)\in \ \mathcal{E}_{\pi}$, this can be guaranteed by imposing appropriate constraints over the parameter vectors $\vec{\eta}_{rs}^l,\; \forall l\in \bar{\mathcal{K}}_{rs}$ according to the following result.
\begin{proposition}\label{communicaiton resolution}
    Consider an edge $(r,s) \in \mathcal{E}_{\pi}$ and the parametric task $\varphi_{rs}^{l}$ with $l \in \bar{\mathcal{K}}_{rs}$ as specified in \eqref{eq:overloading}. Furthermore, consider the truth set $\mathcal{B}_{rs}^l = \mathcal{P}(A_{rs}^l, \vec{c}_{rs}^{l}, \alpha_{rs}^{l}\vec{z}_{rs}^l)$ and parameter vector $\vec{\eta}_{rs}^{l}$ associated with $\varphi_{rs}^l$. Let the generator set $V_{rs}^l = \nu(\mathcal{P}(A_{rs}^l,\vec{0},\vec{z}_{rs}^l))$ as per Prop. \ref{equivalent represenations}. Then  $\varphi_{rs}^l$ is communication consistent if 
\begin{equation}\label{eq: each vertex norm constrained}
(\vec{\eta}_{rs}^{l})^T N^l_k \vec{\eta}_{rs}^{l} \leq r_c^2, \; \forall k= 1,\ldots |V_{rs}^l| ,
\end{equation}
where  
\begin{equation}
\begin{aligned}\label{eq:communication consistency matrices}
N^l_k &= (G^l_k)^T S^TSG^l_k , \; G^l_k &= \comp{I_{n}}{\vec{v}_k},\; \vec{v}_k \in V_{rs}^l.
\end{aligned}
\end{equation}
\end{proposition}
\begin{proof} 
We prove that \eqref{eq: each vertex norm constrained} implies  \eqref{eq:communication consistency equation}. First let the polytope  $\mathcal{L}_{rs} = \{S \vec{e}_{rs}=\vec{p}_{s}-\vec{p}_r\;|\; \vec{e}_{rs}\in \mathcal{B}_{rs}^{l}\}$ represent the projection of $\mathcal{B}_{rs}^{l}$ over the space of relative positions of agent $r$ and $s$. Moreover, let the distance function $f(S\vec{e}_{rs}) = \|S\vec{e}_{rs}\|^2 -r^2_c, \forall S\vec{e}_{rs}\in \mathcal{L}_{rs}$. By Def. \ref{convex hull} and \eqref{eq:generators equivalence} the generators of $\mathcal{L}_{rs}$ are given by the vectors $\{SG^l_k\vec{\eta}_{rs}^{l}\}_{k=1}^{|V^l_{rs}|}$, where $\{G^l_k\vec{\eta}_{rs}^{l}\}_{k=1}^{|V^l_{rs}|} = \nu(\bar{\mathcal{B}}_{rs}^{l}) = \nu(\mathcal{P}(A_{rs}^l,\vec{c}_{rs}^{l},\alpha_{rs}^{l}\vec{z}_{rs}^l))$. Thus  $S\vec{e}_{rs} = \sum_{k} \lambda_k S G^l_k\vec{\eta}_{rs}^{l}, \; \forall S\vec{e}_{rs} \in  \mathcal{L}_{rs}$ with $\lambda_k>0$,   $\sum_{k} \lambda_k=1$. From, Jensen's inequality \cite[Thm. 3.4]{rockafellar1997convex} we also have that $f(SG^l_k\vec{\eta}_{rs}^{l})\leq 0 ,\; \forall k = 1,\ldots |V_{rs}^l| \Rightarrow f(S\vec{e}_{rs})=\|S\vec{e}_{rs}\|^2 -r_c^2 = \|\vec{p}_r- \vec{p}_s\|^2 -r_c^2 \leq 0, \; \forall S\vec{e}_{rs} \in  \mathcal{L}_{rs}$. It is then sufficient to note that $f(SG^l_k\vec{\eta}_{rs}^{l})\leq 0 ,\; \forall k = 1,\ldots |V_{rs}^l|$ is equivalent to \eqref{eq: each vertex norm constrained} such that $\mathcal{B}_{rs}^{l} \cap \{\vec{e}_{rs} \in \mathbb{X}_{rs}\; |\; \|S\vec{e}_{rs}\| \leq r_c \} = \mathcal{L}_{rs}$, which implies  \eqref{eq:communication consistency equation}.
\end{proof}

Note that the matrices $N^l_k$ are positive semi-definite such that relations \eqref{eq: each vertex norm constrained} are convex in $\vec{\eta}_{rs}^{l}$.

\section{Parameters optimization}\label{optimization section}
In Sec. \ref{task decomposition} we have developed a framework to decompose communication inconsistent collaborative tasks in the form $\phi_{ij} =\varphi_{ij}$ by introducing parametric tasks within the family of parametric predicates in \eqref{eq:parameteric predicate}. Furthermore, we have provided relevant constraints on the parameters of the parametric tasks to ensure conflict-free conjunctions of tasks over the communication consistent edges $\mathcal{E}_{\pi}$ applied for the decomposition and to ensure communication consistency of these. In this section, we show how convex optimization can be leveraged to obtain optimal parameters $\vec{\eta}_{rs}^{l},\; \forall (r,s)\in \mathcal{E}_{\pi}, \; \forall l\in \bar{\mathcal{K}}_{rs}$ that satisfy the aforementioned constraints and such that the \textit{decomposition accuracy} as \eqref{eq:task decomposition accuracy} is maximised.
\subsection{Centralized parameters optimization}\label{centralized optimization program}
The convex optimization program to solve the decomposition of the inconsistent tasks is presented next by defining the variables, cost function and constraints.

\subsubsection{Variables} The variables of the optimization program are the parameters $\vec{\eta}_{rs}^{l},\; \forall (r,s)\in \mathcal{E}_{\pi}$ of the tasks $\varphi_{rs}^l$ introduces by the decomposition for all $l\in \bar{\mathcal{K}}_{rs}$ and all $(r,s) \in \mathcal{E}_{\pi}$, together with all the auxiliary variables $\vec{\xi}_{rs}^{q}\in \Xi_{rs},\; \forall (r,s)\in \mathcal{E}_{\pi},\; \forall q =1, \ldots |\Xi_{rs}|$ (as per \eqref{eq:auxiliary set}) applied to avoid conflicting conjunctions as explained in Sec. \ref{handling communication}. We thus define the set of parameters $\mathcal{H}_{rs}$ for each edge $(r,s)$ as $\mathcal{H}_{rs} = \{ \vec{\eta}_{rs}^{l}\, | \, l \in  \bar{\mathcal{K}}_{rs} \}$. We recall that the mapping $y_{\Pi_{rs}}$ as per \eqref{eq:edge mapping for new tasks} assigns a unique index $l\in \bar{\mathcal{K}}_{rs}$ to each inconsistent edge $(i,j) \in \Pi_{rs}$ passing though $(r,s)$. Indeed,  when an inconsistent task $\varphi_{ij}$ with decomposition path  $\vec{\pi}_i^j$ passes through $(r,s)$, then a collaborative task $\bar{\varphi}_{rs}^{ij} \underset{y_{\Pi_{rs}}}{=} \varphi_{rs}^{l}$, with $l\in \bar{\mathcal{K}}_{rs}$, is introduced over $(r,s)$.
\subsubsection{Cost Function}
Let a single inconsistent collaborative task $\phi_{ij} = \varphi_{ij}$ with truth set $\mathcal{B}_{ij}$ be decomposed over the path $\vec{\pi}_{i}^{j}$. By Lemma \ref{single formula decomposition}, the inclusion relation  \eqref{eq:minkosky inclusion} must hold for a valid decomposition, and by Prop. \ref{inclusion accuracy} we already argued that the satisfaction of \eqref{eq:minkosky inclusion} implies the relation \eqref{eq:task decomposition accuracy} $\sum_{(r,s) \in \epsilon(\vec{\pi}_i^j)} \alpha_{rs}^{ij} \leq 1$ (where the notation in place before the definition of the index mapping $y_{\Pi_{rs}}$ is used here and replaced in \eqref{eq:const equivalence}), which holds with equality if and only if \eqref{eq:minkosky inclusion} does.  Intuitively, the optimal decomposition is indeed obtained when \eqref{eq:minkosky inclusion} holds with equality as this is equivalent to recovering the truth set $\mathcal{B}_{ij}$ in its full extent, after the decomposition. Since this condition occurs only when $\sum_{(r,s) \in \epsilon(\vec{\pi}_i^j)} \alpha_{rs}^{ij} = 1$, then the decomposition accuracy in \eqref{eq:task decomposition accuracy} is considered further as an appropriate metric to define "optimality" for our proposed decomposition. Therefore, the objective function of our decomposition becomes 
\begin{equation} \label{eq:const equivalence}
\begin{aligned}
 \sum_{(i,j) \in \mathcal{E}_{\psi} \setminus \mathcal{E}_{c}} \sum_{(r,s) \in \epsilon(\vec{\pi}_i^j)} \alpha_{rs}^{ij} =   \sum_{(r,s) \in \mathcal{E}_{\pi}} \sum_{(i,j) \in \Pi_{rs}} \alpha_{rs}^{ij}  \\ =  \sum_{(r,s) \in \mathcal{E}_{\pi}} \sum_{l \in \bar{\mathcal{K}}_{rs}} \alpha_{rs}^{l},
 \end{aligned}
\end{equation}
where the first equality is obtained by swapping an iteration over the inconsistent edges $(i,j) \in \mathcal{E}_{\psi} \setminus \mathcal{E}_{c}$, with an iteration over the decomposition edges $(r,s) \in \mathcal{E}_{\pi}$. On the other hand, the last equality is only notational and obtained by leveraging the index map $y_{\Pi_{rs}}: \Pi_{rs} \mapsto \bar{\mathcal{K}}_{rs}$ as per \eqref{eq:edge mapping for new tasks}.

\subsubsection{Constraints}
The constraints for our optimization program are of three main types: communication consistency, conflicting conjunctions and decomposition constraints. Consider first the communication constraints over an edge $(r,s) \in \mathcal{E}_{\pi}$. For each parametric task $\varphi_{rs}^l, \;l\in \bar{\mathcal{K}}_{rs}$ there is a parametric truth set $\mathcal{B}_{rs}^l = \mathcal{P}(A_{rs}^l,\vec{c}_{rs}^l, \alpha_{rs}^l \vec{z}_{rs}^l)$ with generators set $V_{rs}^l = \nu(\mathcal{P}(A_{rs}^l,\vec{0},\vec{z}_{rs}^l))$. Then, by Prop. \ref{communicaiton resolution}, communication consistency is enforced by \eqref{eq: each vertex norm constrained}. We can repeat the same argument for each $l\in \bar{\mathcal{K}}_{rs}$ to obtain :
$$
(\vec{\eta}_{rs}^l )^T N^l_{k}\vec{\eta}_{rs}^l \leq r_c^2,\; \forall l \in \bar{\mathcal{K}}_{rs},\; \forall k=1, \ldots  |V_{rs}^l|
$$
where the matrices $N_k^l, \; \forall k =1,\ldots |V_{rs}^l|$ are as per \eqref{eq:communication consistency matrices}.  \par 
The second type of constraint guarantees conflict-free conjunctions over the tasks $\phi_{rs}$ as per \eqref{eq:overloading}. In this respect, consider, for each edge $(r,s) \in \mathcal{E}_{\pi}$, the set $\mathfrak{Q}_{rs}$ as per \eqref{eq:the set of indices for conflicts}. The set $\mathfrak{Q}_{rs}$ is the set of subsets of indices in $\mathcal{K}_{rs}$ such that, for each $Q\in \mathfrak{Q}_{rs}$, a conjunction of tasks $\land_{l\in Q} \varphi_{rs}^l$ over $(r,s)\in \mathcal{E}_{\pi}$ could cause a conflict.  Then, conflicting conjunctions are avoided by imposing \eqref{eq:elegant conflict resolution equation} over each edge $(r,s) \in \mathcal{E}_{\pi}$. \par
The third, and last, type of constraint is the one required to obtain a valid decomposition of each inconsistent task $\phi_{ij}= \varphi_{ij},\; \forall (i,j) \in \mathcal{E}_{\psi} \setminus \mathcal{E}_c$ by the satisfaction of \eqref{eq: linear inclusion}. This is achieved by the satisfaction of \eqref{eq: linear inclusion} and \eqref{eq:definition of the intersection inclusion matrix}. Differently from the previous two types of constraints, this constraint is \textit{shared} among multiple edges $(r,s) \in \epsilon(\vec{\pi}_{i}^{j})$ in the sense that its satisfaction involves parameters vectors spanning all the edges $(r,s) \in \epsilon(\vec{\pi}_{i}^{j})$ along the path $\vec{\pi}_{i}^{j}$. With the intent of uniforming the notation via the index mapping $y_{\Pi_{rs}}$, we let $l_{rs} = y_{\Pi_{rs}}((i,j))$ for each $(r,s)\in \epsilon(\vec{\pi}_i^j)$ to be the unique index associated with the inconsistent edge $(i,j)$ over the edge $(r,s)\in \epsilon(\vec{\pi}_i^j)$, such that the constraint \eqref{eq: linear inclusion} can be rewritten as 
$$
\sum_{(r,s)\in \epsilon(\vec{\pi}_i^j)} \Biggl( M_{ij}\vec{\eta}^{l_{rs}}_{rs} - \frac{Z_{ij}}{|\epsilon(\vec{\pi}_i^j)|}\begin{bmatrix}\vec{c}_{ij}\\1\end{bmatrix} \Biggr) \leq \vec{0},
$$
 for all $(i,j) \in \mathcal{E}_{\psi} \setminus \mathcal{E}_{c}$. Recalling that the truth set of the inconsistent task $\varphi_{ij}$ is given by $\mathcal{B}_{ij} = \mathcal{P}(A_{ij}, \vec{c}_{ij}, \vec{z}_{ij})$, the definition of $M_{ij}$ and $Z_{ij}$ is given in \eqref{eq:definition of the intersection inclusion matrix}.

\subsubsection{Optimization program}
By letting $\mathcal{H}=\bigcup_{(r,s) \in \epsilon(\vec{\omega})} \mathcal{H}_{rs}$ and $\Xi = \bigcup_{(r,s) \in \epsilon(\vec{\omega})} \Xi_{rs}$ the centralised optimization program applied to solve the task decomposition problem is then given by:
\begin{subequations}\label{eq:convex optimization problem}
\begin{equation}\label{eq:centralised cost}
\max_{\mathcal{H}, \Xi} \; \sum_{(r,s) \in \mathcal{E}_{\pi}} \sum_{l \in \bar{\mathcal{K}}_{rs}} \alpha_{rs}^{l}
\end{equation}
\begin{equation} \label{eq:communication constraint}
(\vec{\eta}_{rs}^{l})^T N^l_k (\vec{\eta}_{rs}^{l}) \leq r^2_c
\begin{array}{c}
\forall (r,s) \in \mathcal{E}_{\pi},\\
\forall l \in \bar{\mathcal{K}}_{rs},\\
\qquad \qquad  \forall k =1, \ldots |V^l_{rs}|, \\
\end{array}
\end{equation}
\begin{equation} \label{eq:all inclusion constraints}
\begin{array}{c}
 A_{rs}^l \vec{\xi}_{rs}^q + \comp{A_{rs}^l}{\vec{z}_{rs}^l} \vec{\eta}_{rs}^{l} \leq \vec{0}, \quad  \\
 \text{with}\;  q=y_{\mathfrak{Q}_{rs}}(Q)
 \end{array}
 \begin{array}{cc}
\forall (r,s) \in \mathcal{E}_{\pi},\\
\forall Q \in \mathfrak{Q}_{rs},\\
\forall l \in Q, \\
\end{array}
\end{equation}

\begin{equation}\label{eq: implication constraint}
\begin{array}{c}
\sum_{(r,s)\in \epsilon(\vec{\pi}_i^j)} \Biggl( M_{ij}\vec{\eta}^{l_{rs}}_{rs} - \frac{Z_{ij}}{|\epsilon(\vec{\pi}_i^j)|} ,\begin{bmatrix}\vec{c}_{ij}\\1\end{bmatrix} \Biggr) \geq 0,\\
\text{with} \; l_{rs} = y_{\Pi_{rs}}((i,j)), \qquad \qquad  \forall (i,j)\in \mathcal{E}_\psi\setminus \mathcal{E}_c,
\end{array}
\end{equation}
\begin{equation}\label{eq: norm bound}
\begin{array}{ll}
    \|\vec{\xi}_{rs}^q\| \leq \xi_{max}    &\forall q =1, \ldots |\mathfrak{Q}_{rs}|,\\
    \|\vec{\eta}_{rs}^{l}\| \leq \eta_{max} &\forall l\in \bar{\mathcal{K}}_{rs}, 
\end{array}
\quad \forall (r,s) \in \mathcal{E}_{\pi}
\end{equation}
\end{subequations}
where the cost function \eqref{eq:centralised cost} represents the sum of all the decomposition accuracies over all the inconsistent tasks, constraint \eqref{eq:communication constraint} enforces communication consistency over all parametric tasks introduced for the decomposition, constraint \eqref{eq:all inclusion constraints} enforces all the necessary inclusions to avoid conflicting conjunctions over the decomposition edges in $\mathcal{E}_{\pi}$ and constraint \eqref{eq: implication constraint} represents all the inclusion relations needed to satisfy \eqref{eq:minkosky inclusion} for all the inconsistent tasks, with $M_{ij}$ and $Z_{ij}$ defined as per \eqref{eq:definition of the intersection inclusion matrix} for each $(i,j) \in \mathcal{E}_{\psi}\setminus \mathcal{E}_{c}$. Constraint \eqref{eq: norm bound} with constants $\xi_{max}>0$ and $\eta_{max}>0$, is a norm constraint introduced to ensure the domain of the convex program \eqref{eq:convex optimization problem} is compact. The upper bounds $\xi_{max}$, $\eta_{max}$ should be chosen to be sufficiently large as they only apply to bound the domain of the optimal parameters search. 

\subsection{A decentralized solution}\label{distributed task decompostion section}
While problem \eqref{eq:convex optimization problem} can be solved in a centralised fashion, the next proposition confirms the intuition that \eqref{eq:convex optimization problem} can be written as a set of constraint-coupled programs defined over the edges $(r,s) \in \mathcal{E}_{\pi}$, which are coupled via the shared constraint \eqref{eq: implication constraint}.
\begin{proposition}\label{eq:equivalent program proposition}
For each $(r,s)\in \mathcal{E}_{\pi}$, let the stacked vector of parameters and auxiliary variables $\vec{\chi}_{rs} \in \mathbb{R}^{(n+1)\cdot|\mathcal{H}_{rs}| + n\cdot|\mathfrak{Q}_{rs}|}$ defined as
\begin{equation}
\vec{\chi}_{rs} = [ \; [\vec{\eta}_{rs}^{l}]_{l \in \bar{\mathcal{K}}_{rs}},\; [\vec{\xi}_{rs}^q]_{q=1,\ldots |\mathfrak{Q}_{rs}|} \;].
\end{equation}
Then the optimization program \eqref{eq:convex optimization problem} can be rewritten as
\begin{subequations}\label{eq:structured program}
\begin{align}
\min_{\vec{\chi}_{rs} \forall (r,s) \in \mathcal{E}_{\pi}} &\sum_{(r,s)\in \mathcal{E}_{\pi}} f_{rs}(\vec{\chi}_{rs})\label{eq:compact cost function}\\
\text{s.t. : } \; \vec{\chi}_{rs} &\in \mathcal{C}_{rs} \qquad \forall (r,s) \in \mathcal{E}_{\pi} \label{eq:compact convex sets}, \\
\sum_{(r,s) \in \mathcal{E}_{\pi}} ( T_{rs} \vec{\chi}_{rs}& - \vec{t}_{rs})  \leq \vec{0} \label{eq:compact shared constraint}, 
\end{align}
\end{subequations}
where the functions\footnote{The functions $f_{rs}$ should not be confused with the system dynamics as it is clear from the context} $f_{rs}(\vec{\chi}_{rs}) = \sum_{l\in \mathcal{K}_{rs}} \alpha_{rs}^l,\; \forall (r,s)\in \mathcal{E}_{\pi} , \; \forall (r,s) \in \mathcal{E}_{\pi}$ are convex and the sets $\mathcal{C}_{rs}, \; \forall (r,s) \in \mathcal{E}_{\pi}$ are compact and convex sets. Moreover, $T_{rs} \in \mathbb{R}^{(\sum_{(i,j) \in  \mathcal{E}_{\psi} \setminus \mathcal{E}_{c}}m_{ij} ) \times ((n+1)\cdot|\mathcal{H}_{rs}| + n\cdot|\Xi_{rs}|)}$ and $\vec{t}_{rs} \in \mathbb{R}^{(\sum_{(i,j) \in  \mathcal{E}_{\psi} \setminus \mathcal{E}_{c}}m_{ij} )}$ with $m_{ij}$ is the number of rows of each matrix $M_{ij}$ as per \eqref{eq:definition of the intersection inclusion matrix}.
\end{proposition}
\begin{proof}
First note that $f_{rs}(\vec{\chi}_{rs}), \forall (r,s)\in \mathcal{E}_{\Pi}$, are linear (and thus convex) functions since they are obtained as sums of elements of the vectors $\vec{\chi}_{rs}$. Next, we prove 1) the equivalence between \eqref{eq:centralised cost} and \eqref{eq:compact cost function}, 2) the equivalence between \eqref{eq:communication constraint}-\eqref{eq:all inclusion constraints}-\eqref{eq: norm bound} and  \eqref{eq:compact convex sets}, 3) the equivalence between \eqref{eq: implication constraint} and \eqref{eq:compact shared constraint}. The equivalence 1) of the cost function \eqref{eq:compact cost function} and \eqref{eq:centralised cost} is evident since $f_{rs}(\vec{\chi}_{rs}) = \sum_{l\in \mathcal{K}_{rs}} \alpha_{rs}^l$. The equivalence 2) follows by considering, for each $(r,s) \in \mathcal{E}_{rs}$, the sets  $\mathcal{C}^1_{rs} = \{ \vec{\chi}_{rs}\,|\, (\vec{\eta}_{rs}^{l})^T N^l_k (\vec{\eta}_{rs}^{l}) \leq r^2_c, \forall l\in \bar{\mathcal{K}}_{rs}\; k = 1, \ldots |V_{rs}^l|\}
$, 
$\mathcal{C}^2_{rs}= \{ \vec{\chi}_{rs}\,|\,  A_{rs}^l \vec{\xi}_{rs}^{q} + \comp{A_{rs}^l}{\vec{z}_{rs}^l} \vec{\eta}_{rs}^{l} \leq \vec{0},\; \forall l \in \bar{\mathcal{K}}_{rs}, \forall Q\in \mathfrak{Q}_{rs},\;  \text{s.t},\; q = y_{\mathfrak{Q}_{rs}(Q)} \}
$, 
$
\mathcal{C}^3_{rs}= \{ \vec{\chi}_{rs}\,|\, \|\vec{\xi}_{rs}^q\| \leq \xi_{max}\; \forall q= 1,\ldots |\mathfrak{Q}_{rs}| \land \|\vec{\eta}_{rs}^{l}\| \leq \eta_{max}\; \forall l \in \bar{\mathcal{K}}_{rs}\},
$
which are equivalent to \eqref{eq:communication constraint}-\eqref{eq:all inclusion constraints}-\eqref{eq: norm bound} when considered for a single edge $(r,s) \in \mathcal{E}_{\pi}$. Since  $\mathcal{C}^1_{rs}$, $\mathcal{C}^2_{rs}$ and $\mathcal{C}^3_{rs}$ are convex, and $\mathcal{C}^3_{rs}$ is bounded, then $\mathcal{C}_{rs} = \mathcal{C}^1_{rs} \cap \mathcal{C}^2_{rs} \cap \mathcal{C}^3_{rs}$ is also convex and bounded, thus proving the equivalence. The last equivalence between \eqref{eq: implication constraint} and \eqref{eq:compact shared constraint} is proved as follows. Consider, for each $(r,s) \in \mathcal{E}_{\pi}$, the matrix $W_{rs} \in  \mathbb{R}^{(\sum_{(i,j) \in  \mathcal{E}_{\psi} \setminus \mathcal{E}_{c}}m_{ij} ) \times (n+1)\cdot|\Pi_{rs}|}$, where we recall that $\Pi_{rs}$ contains all the inconsistent edges $(i,j)\in \mathcal{E}_{\psi} \setminus \mathcal{E}_{c}$ whose decomposition path is passing by $(r,s)\in \mathcal{E}_{\pi}$ as per \eqref{eq:definition of pi edge} and $|\Pi_{rs}| = |\bar{\mathcal{K}}_{rs}|$ by \eqref{eq:parametric set of indices}. Moreover $m_{ij}$ is the number of rows of the matrix $M_{ij}$ associated with each inconsistent each edge $(i,j) \in \mathcal{E}_{\psi} \setminus \mathcal{E}_{c}$. We define the matrices $W_{rs}$ by blocks as $W_{rs}[p,w]$ with row-blocks $p = 1,\ldots |\mathcal{E}_{\psi}\setminus\mathcal{E}_{c}|$ and $w = 1, \ldots |\Pi_{rs}|$ column-blocks such that all the matrices $W_{rs}$ have the same number of row-blocks, but only a number of column-blocks equal to the number of edges $|\Pi_{rs}|$ that are decomposed over the edge $(r,s) \in \mathcal{E}_{\pi}$ (and recall that $|\bar{\mathcal{K}}_{rs}|=|\Pi_{rs}|$). We define the one-to-one index mapping $y_{\mathcal{E}_{\psi}\setminus\mathcal{E}_{c}} :  \mathcal{E}_{\psi}\setminus\mathcal{E}_{c} \rightarrow \{1, \ldots |\mathcal{E}_{\psi}\setminus\mathcal{E}_{c}|\}$ associating each inconsistent edges $(i,j) \in \mathcal{E}_{\psi} \setminus \mathcal{E}_c$ with a unique row-block index. Then each matrix $W_{rs}$ has non-zero blocks only at those $(p,w)$ index pairs such that $(i,j) =  y_{\mathcal{E}_{\psi} \setminus \mathcal{E}_c}^{-1}(p)$ is an inconsistent edge and the mapping $y_{\Pi_{rs} }^{-1}(w + K_{rs}) = (i,j)$, as per \eqref{eq:edge mapping for new tasks}, is well defined, meaning that the decomposition path $(i,j) \in \mathcal{E}_{\psi} \setminus \mathcal{E}_c$ passes though $(r,s)$. Thus, $W_{rs}$ is defined by blocks as
$$ 
W_{rs}[p,w] = \begin{cases} 
M_{ij}  \quad \text{if} \,  
      y_{\mathcal{E}_{\psi} \setminus \mathcal{E}_c}^{-1}(p) = y_{\Pi_{rs} }^{-1}(w + K_{rs}) = (i,j)  \\
\vec{0}_{m_{ij}\times (n+1)} \quad \text{else},
\end{cases}
$$
At this point we define the selection matrix $S_{rs}$ of appropriate dimensions such that $[\vec{\eta}_{rs}^l]_{l\in \bar{\mathcal{K}}_{rs}} = S_{rs}\vec{\chi}_{rs}$ and finally $T_{rs} = W_{rs}S_{rs}$. Concerning to vector $\vec{t}_{rs}$, this can also be defined by a number of $|\mathcal{E}_{\psi} \setminus \mathcal{E}_{c}|$ row-blocks of dimension $\mathbb{R}^{m_{ij}}$ as
$$
\vec{t}_{rs}[p] = \begin{cases} 
\frac{Z_{ij}}{|\epsilon(\vec{\pi}_i^j)|} \begin{bmatrix}
    \scriptstyle\vec{c}_{ij}\\ \scriptstyle 1
\end{bmatrix}\quad \text{if} \,  
      y_{\mathcal{E}_{\psi} \setminus \mathcal{E}_c}^{-1}(p)=(i,j) \in \Pi_{rs}  \\
\vec{0}_{m_{ij}} \hspace{1.5cm} \text{else},
\end{cases}
$$
With these definitions, if we let $\vec{o}_{rs} = T_{rs} \vec{\chi}_{rs} - \vec{t}_{rs}$, then $\vec{o}_{rs}$ is a vector defined by a number of $|\mathcal{E}_{\psi} \setminus \mathcal{E}_{c}|$ blocks of dimension $\mathbb{R}^{m_{ij}}$ such that 
\begin{equation}\label{eq:sparsity pattern}
\vec{o}_{rs}[p] = \begin{cases} 
\begin{array}{c}
      M_{ij} \vec{\eta}_{rs}^{l_{rs}} - \frac{Z_{ij}}{|\epsilon(\vec{\pi}_i^j)|} \begin{bmatrix}
    \scriptstyle\vec{c}_{ij}\\ \scriptstyle 1\\
\end{bmatrix}\\
     \text{s.t}\;  l_{rs} = y_{\Pi_{rs}}((i,j))
\end{array}
\begin{array}{l}
      \,\text{if} \,  y_{\mathcal{E}_{\psi} \setminus \mathcal{E}_c}^{-1}(p) \\
      \quad =(i,j) \in \Pi_{rs} 
\end{array}\\\\
\begin{array}{c}
\vec{0}_{m_{ij}} \hspace{3.cm}\text{else},
\end{array}
\end{cases}
\end{equation}
where the notation $l_{rs}$ is introduced again to be consistent with \eqref{eq: implication constraint} and make clear that $l_{rs} = y_{\Pi_{rs}}((i,j))$ is the index assigned to the inconsistent edge $(i,j)$ over $(r,s) \in \epsilon(\vec{\pi}_i^j)$, which is different from the index $l_{r's'} = y_{\Pi_{r's'}}((i,j))$ assigned to the same edge $(i,j)$ by another edge $(r',s') \in \epsilon(\vec{\pi}_i^j)$. Eventually, noting that $\sum_{(r,s) \in \mathcal{E}_{\pi}} \vec{o}_{rs}$ is equivalent to \eqref{eq: implication constraint} concludes the proof.
\end{proof}
\par 

Given the optimization problem \eqref{eq:convex optimization problem}, a solution can be found in a decentralized fashion by leveraging the fact that \eqref{eq:convex optimization problem} has the edge-wise decentralized structure in \eqref{eq:structured program}. We propose to employ the algorithm in \cite{notarnicola2019constraint} to solve \eqref{eq:structured program} in a decentralized fashion as explained next. First, we wish to define an edge-computing graph denoted as $\mathcal{G}_{\Theta}$ such that each node in $\mathcal{G}_{\Theta}$ corresponds to an edge $(r,s) \in \mathcal{E}_{\pi}$. Hence, the set of nodes for $\mathcal{G}_{\Theta}$ is given by $\mathcal{V}_{\theta} = \mathcal{E}_{\pi}$. On the other hand, the set of edges for $\mathcal{G}_{\Theta}$ is given by $\mathcal{E}_{\Theta} = \{ ((r,s),(r',s')) \in \mathcal{V}_{\Theta} \times \mathcal{V}_{\Theta} \; | \;  \{r,s\} \cap \{r',s'\} \neq \emptyset \}$. In other words, a tuple $((r,s),(r',s'))$ is an edge in $\mathcal{E}_{\Theta}$ if $(r,s)\in \mathcal{E}_{\pi}$ shares a node with  $(r',s') \in \mathcal{E}_{\pi}$. Then the edge-computing graph is defined as $\mathcal{G}_{\Theta}(V_{\Theta},\mathcal{E}_{\Theta})$ and the neighbour set is defined as $\mathcal{N}_{\Theta}((r,s)) = \{ 
(r',s') \in \mathcal{V}_{\Theta} |  ((r,s),(r',s')) \in \mathcal{E}_{\Theta} \}$. The edge-computing graph $\mathcal{G}_{\Theta}$ can then be applied to find a solution for \eqref{eq:structured program} in a decentralized fashion where each node $(r,s) \in \mathcal{V}_{\Theta}$ (an edge in $\mathcal{E}_{\pi}$) solves the following relaxed local problem iteratively 
\begin{subequations}\label{eq:distributed opt}
\begin{align}
&\min_{\vec{\chi}_{rs}, \rho_{rs}}  f_{rs}(\vec{\chi}_{rs}) +  c_{\rho}\rho_{rs},       \label{eq:compact cost function single}\\
&\text{s.t. : }\; \vec{\chi}_{rs} \in \mathcal{C}_{rs}, \; \rho_{rs} \geq 0 , \label{eq:compact convex sets single} \\
T_{rs} \vec{\chi}_{rs} - \vec{t}_{rs} +  &\sum_{(r's')\in \mathcal{N}_{\Theta}((r,s))} ({}^{t}\lambda_{rs,r's'} - {}^{t}\lambda_{r's',rs}) \leq \vec{1}\rho_{rs}.\label{eq:compact shared constraint single} 
\end{align}
\end{subequations}
In  \eqref{eq:distributed opt}, the superscript $t$ indicates the current iteration and is set on the upper-left to avoid confusion with the previous notation. The term $\rho_{rs}\in \mathbb{R}$ is a penalty variable constrained to be non-negative as per constraint \eqref{eq:compact convex sets single}. The coefficient $c_{\rho} >0$ penalizes non-zero values of $\rho_{rs}$. At the same time, by \eqref{eq:compact shared constraint single}, each node $(r,s)$ holds the value at iteration $t$ of the additional \textit{consensus} vectors ${}{}^{t}\vec{\lambda}_{rs,r's'},  \in \mathbb{R}^{(\sum_{(i,j) \in  \mathcal{E}_{\psi} \setminus \mathcal{E}_{c}}m_{ij})}, \; \forall (r',s')\in \mathcal{N}_{\Theta}((r,s))$ and receives the consensus vectors ${}{}^{t}\vec{\lambda}_{r's',r,s}$ from the each neighbour $(r',s')\in \mathcal{N}_{\Theta}((r,s))$. The introduction of a consensus term like the one introduced in \eqref{eq:distributed opt} is common in distributed optimization of a global optimization program with a separable structure as the one in \eqref{eq:structured program} \cite{notarnicola2019constraint,falsone2017dual}. Note that summing the left-hand side of constraint \eqref{eq:compact shared constraint single} for each $(r,s)\in \mathcal{E}_{\pi}= \mathcal{V}_{\Theta}$ yields the original left-hand side in \eqref{eq:compact shared constraint}. At this point  Algorithm \ref{alg:rsdd} borrowed from \cite{notarnicola2019constraint} applies to converge to an optimal solution of the vectors $\vec{\chi}_{rs}, \; \forall (r,s) \in \mathcal{E}_{\pi}$, after sufficiently many iterations. While referring the reader to \cite{notarnicola2019constraint} for a detailed analysis of Alg. \ref{alg:rsdd}, we here seek to provide an intuitive explanation of the algorithm. Namely, at each iteration $t$, the edge $(r,s) \in \mathcal{V}_{\Theta}$ solves an instance of \eqref{eq:distributed opt}, where the penalty variable $\rho_{rs}\in \mathbb{R}$ relaxes the satisfaction of the shared constraint \eqref{eq:compact shared constraint single}. If we let ${}^{t}\rho_{rs}$ to be the optimal value of $\rho_{rs}$ at iteration $t$ and if we let ${}^{t}\vec{\mu}_{rs} \in \mathbb{R}^{\sum_{(i,j) \in \mathcal{E}_{\psi} \setminus \mathcal{E}_{c}}m_{ij}}$ to be the Lagrangian multiplier vector associated with \eqref{eq:compact shared constraint single}  at each iteration $t$, then the consensus vectors $\mathcal{\lambda}_{rs,r's'}$ are updated by node $(r,s)$, at iteration $t$, as per line \ref{lst:line:update} in Alg, \ref{alg:rsdd}. Note that this update only requires the local knowledge of the Lagrangian multiplier vectors $\vec{\mu}_{rs}$, $\vec{\mu}_{r's'}, \; \forall (r',s') \in \mathcal{N}_{\Theta}((r,s))$. The update rate ${}^{t}\gamma>0$ is must be known to all agents and such that $\lim_{t\rightarrow \infty}{}^{t}\gamma = \infty$ and $\lim_{t\rightarrow \infty}({}^{t}\gamma)^2 < \infty$ to guarantee convergence \cite{notarnicola2019constraint}. If we let $[\vec{\chi}_{rs}^{\star}]_{(r,s) \in \mathcal{E}_{\pi}}$ to be a non-unique optimal (feasible) solution of \eqref{eq:structured program}, then, by \cite[Thm II.6]{notarnicola2019constraint}, it is known that the iterates of optimal solutions $\{[{}^{t}\vec{\chi}_{rs}]_{\forall (r,s)\in \mathcal{E}_{\pi}}\}$ found by Alg. \ref{alg:rsdd} converge asymptotically to one such solution $[\vec{\chi}_{rs}^{\star}]_{(r,s) \in \mathcal{E}_{\pi}}$. This, in turns, entails that $\lim_{t\rightarrow \infty} {}^t\rho_{rs} = 0,\; \forall (r,s)\in \mathcal{V}_{\Theta}$. Once Alg. \ref{alg:rsdd} terminates for all the agents, a set of optimal parameters $(\vec{\eta}_{rs}^l)^{\star}\; \forall l\in \bar{\mathcal{K}}_{rs}$, is obtained from each edge $(r,s) \in \mathcal{E}_{\pi}$ such that the parametric tasks $\varphi_{rs}^l \underset{y_{\Pi_{rs}}}{=} \bar{\varphi}_{rs}^{ij}$ as per \eqref{eq:specific form of phi bar} are fully determined as well as the new global task $\bar{\psi}$ as per \eqref{eq:global specification rewritten}, thus solving Problem \ref{problem1}.
\begin{algorithm}[b]
\caption{ For single edge $(r,s)\in \mathcal{E}_{\pi} = \mathcal{E}_{\Theta}$ \cite{notarnicola2019constraint}}\label{alg:rsdd}
\begin{algorithmic}[1]
\State \textbf{allocate} $\vec{\chi}_{rs}, \rho_{rs}, \boldsymbol{\mu}_{rs}$,$\vec{\lambda}_{rs,r's'}, \;  \forall (r',s') \in \mathcal{N}_{\Theta}((r,s))$
\State \textbf{initialize} ${}^{0}\vec{\lambda}_{rs,r's'}$ arbitrarily $\forall (r's') \in \mathcal{N}_{\Theta}((r,s))$ and $c_{\rho}>>0$
\While{Not converged}
    \State \textbf{gather} ${}^{t}\vec{\lambda}_{rs,r's'}$ from $(r',s') \in \mathcal{N}_{\Theta}((r,s))$
    \State \textbf{compute} $\left(\left({}^{t+1}\vec{\chi}_{rs}, {}^{t+1}\rho_{rs}\right), {}^{t+1}\vec{\mu}_{rs}\right)$ as primal-dual \newline \hspace*{5.2cm}  optimal of \eqref{eq:distributed opt}
    \For{$(r',s') \in \mathcal{N}_{\Theta}((r,s))$} 
    \State \textbf{gather}  ${}^{t+1}\boldsymbol{\mu}_{r's'}$ and \textbf{update} ${}^{t+1}\lambda_{rs,r's'}$ as \newline  
\hspace*{0.8cm}$
{}^{t+1}\vec{\lambda}_{rs,r's'}={}^{t}\vec{\lambda}_{rs,r's'}- {}^{t}\gamma  \left({}^{t+1}\boldsymbol{\mu}_{rs}- {}^{t+1}\boldsymbol{\mu}_{r's'}\right) 
$\label{lst:line:update}
\EndFor
\State $t = t+1$
\EndWhile
\end{algorithmic}
\end{algorithm}
With these considerations, the final main result of our decomposition is presented next.
\begin{theorem}\label{the last fermat theorem}
     Let $\mathcal{G}_c$ be acyclic as per Assumption \ref{acyclic communication} and let Assumption \ref{conflict free assumption} hold over $\mathcal{G}_{\psi}$. Moreover, let the optimization program \eqref{eq:convex optimization problem} be feasible and solved in a decentralized fashion by the edge-computing graph $\mathcal{G}_{\Theta}$ as per Alg. \ref{alg:rsdd} such that the final global task $\bar{\psi}$ is obtained, together with the task graph $\mathcal{G}_{\bar{\psi}}$, as per \eqref{eq:global specification rewritten}. Then the task graph $\mathcal{G}_{\bar{\psi}}$ is communication consistent and $\bar{\psi}$ does not suffer from conflicting conjunctions as per Fact \ref{conflict 1}-\ref{conflict 5}. Moreover, for all state signals $\vec{x}(t)\in \mathcal{X}$ such that $(\vec{x}(t),0)\models \bar{\psi}$ we have  $(\vec{x}(t),0)\models \psi$.
 \end{theorem}
 \begin{proof}
      When \eqref{eq:convex optimization problem} is feasible, then the satisfaction of the constraints \eqref{eq:communication constraint}-\eqref{eq:all inclusion constraints} implies, by Prop. \ref{conflict resolution 1}, \ref{conflict resolution 2},  \ref{conflict resolution 3} and \ref{communicaiton resolution}, that the resulting tasks $\phi_{ij},\, \forall (i,j) \in \mathcal{E}_{\bar{\psi}}$ are communication consistent and are not conflicting conjunctions. Moreover, by the conditions of Lemma \ref{single formula decomposition}, which are enforced via constraint \eqref{eq: implication constraint}, we know that for each inconsistent task $\phi_{ij}$ with $(i,j)\in \mathcal{E}_{\psi} \setminus \mathcal{E}_{c}$, the conjunction of tasks $\bar{\phi}^{ij} = \land_{(r,s)\in \epsilon(\vec{\pi}_i^j)} \bar{\varphi}_{rs}^{ij}$ as per \eqref{eq:path specification} is such that $(\vec{x}(t),0)\models \bar{\phi}^{ij}\Rightarrow (\vec{x}(t),0)\models \phi_{ij}$. From these considerations, the result of the theorem follows.
 \end{proof}

 \subsection{Computational aspects}
 Concerning some computational aspects, the optimization problem \eqref{eq:convex optimization problem} has a compact convex domain and convex objective such that a bounded set of solutions to \eqref{eq:convex optimization problem} exists if the feasible domain set is not empty. Hence, the proposed decomposition process is sound. The infeasibility of \eqref{eq:convex optimization problem} occurs when the intersection between the domain of \eqref{eq:all inclusion constraints} and \eqref{eq: implication constraint} is empty. This can be intuitively understood as one parametric polytope $\mathcal{B}_{rs}^{l}= \mathcal{P}(A_{rs}^l, \vec{c}_{rs}^{l}, \alpha_{rs}^{l}\vec{z}^l_{rs})$, for some $l\in \bar{\mathcal{K}}_{rs}$, being constrained with an excessive number of intersections that induce a scale factor $\alpha_{rs}^{l}>1$, thus impeding the satisfaction of the shared constraint \eqref{eq: implication constraint} due to the result in Prop. \ref{inclusion accuracy}. To resolve these pathological cases, it is possible to change the structure of the communication topology by leveraging the communication tokens $q_{ij}^c$ hoping that a different selection of decomposition edges $\mathcal{E}_{\pi}$ provides a feasible solution, but we do not provide a direct solution in this direction, which we leave as future work. A few final comments regarding the way information is exchanged over the edge-computing graph $\mathcal{G}_{\Theta}$ and the amount of information required to run Algorithm \ref{alg:rsdd} are given.\par

 First note that that while the edge-computing graph is a convenient abstraction for our decomposition, either agent $r\in \mathcal{V}$ or $s\in \mathcal{V}$ will be effectively solving \eqref{eq:distributed opt}. Since the communication graph $\mathcal{G}_{c}$ is acyclic and $\mathcal{E}_{\pi} \subseteq \mathcal{E}_c$, then it is possible to select the computing nodes by selecting the leaf nodes of $\mathcal{G}_{c}$ as computing and iteratively select computing nodes along the paths from the leaf agent to the root of the graph. In this way the case in which a single agent $r$ solves an instance of \eqref{eq:distributed opt} for both the edges $(r,s')$ and $(r,s'')$ for some $((r,s'), (r,s'')) \in \mathcal{E}_{\Theta}$ is avoided. Moreover, consider again two neighbouring edges $((s',r),(r,s'') \in \mathcal{E}_{\Theta}$ over the computing graph $\mathcal{G}_{\Theta}$ such that the node $r$ is shared. Then exchanging the consensus vectors and Lagrangian parameters ${}^t\vec{\mu}_{rs'},\;{}^t\vec{\mu}_{rs''},\; {}^t\mathcal{\vec{\lambda}}_{rs',rs''}$ and ${}^t\mathcal{\vec{\lambda}}_{rs'',rs'}$ among the neighbouring edges $(s',r)$ and $(r,s'')$ over the communication graph $\mathcal{G}_{c}$, takes either 1 communication hop (if  $s'$ and $r$ are the computing agents for $(s',r)$ and $(r,s'')$, respectively), or 2 communication hops (if $s'$ and $s''$ are the computing agents for the edge $(s',r)$ and $(r,s'')$, respectively).\par

\section{Simulations}\label{simulations}
\begin{table}[]
\centering
\begin{tabular}{ p{1cm} p{1.1cm} p{4.6cm}  }
\toprule[3pt]
Task  &  Operator&  Predicate\\ \midrule
\multicolumn{3}{c}{Exploration task}\\ \midrule
$\phi_1$     & $  G_{[10,20]}$ &$\scriptstyle\langle {}^4A_{1} \left(\vec{x}_1 - [ \scriptscriptstyle5 \; \scriptscriptstyle 0]^T\right) - \vec{1}_5 \cdot 0.4 \rangle $ \\ 
$\phi_{1,8}$ & $F_{[10,15]}$&$\scriptstyle\langle {}^6A_{1,8} \left(\vec{e}_{1,8} - [ \scriptscriptstyle10 \; \scriptscriptstyle 4]^T\right) - \vec{1}_6 \cdot 0.4 \rangle  $ \\ 
$\phi_{1,4}$ & $  F_{[10,15]}$&$\scriptstyle\langle {}^6A_{1,4} \left(\vec{e}_{1,4} - [ \scriptscriptstyle10 \; \scriptscriptstyle -4]^T\right) - \vec{1}_6 \cdot 0.4 \rangle $ \\ 
$\phi_{1,14}$ & $G_{[13,15]}$&$\scriptstyle\langle {}^4A_{1,14} \left(\vec{e}_{1,14} - [ \scriptscriptstyle-12 \; \scriptscriptstyle 5.5]^T\right) - \vec{1}_4 \cdot 0.7 \rangle $ \\ 
$\phi_{1,13}$ & $F_{[10,15]}$&$\scriptstyle\langle {}^5A_{1,13} \left(\vec{e}_{1,13} - [ \scriptscriptstyle-6 \; \scriptscriptstyle -6]^T\right) - \vec{1}_5 \cdot 0.4 \rangle $ \\ 
$\phi_{1,5}$ & $ F_{[13,15]}$&$\scriptstyle\langle {}^5A_{1,5} \left(\vec{e}_{1,5} - [ \scriptscriptstyle-3.5 \; \scriptscriptstyle -4]^T\right) - \vec{1}_5 \cdot 0.4 \rangle $ \\ 
$\phi_{1,6}$ & $G_{[10,15]}$&$\scriptstyle\langle {}^5A_{1,6} \left(\vec{e}_{1,6} - [ \scriptscriptstyle0 \; \scriptscriptstyle 2]^T\right) - \vec{1}_5 \cdot 0.4 \rangle $ \\ 
$\phi_{1,2}$ & $G_{[10,15]}$&$\scriptstyle\langle {}^5A_{1,2} \left(\vec{e}_{1,2} - [ \scriptscriptstyle0 \; \scriptscriptstyle -2]^T\right) - \vec{1}_5 \cdot 0.4 \rangle $ \\ 
$\phi_{1,11}$ & $G_{[10,15]}$&$\scriptstyle\langle {}^5A_{1,11} \left(\vec{e}_{1,11} - [ \scriptscriptstyle-4 \; \scriptscriptstyle 0]^T\right) - \vec{1}_5 \cdot 0.4 \rangle $ \\ 
$\phi_{10,15}$ & $G_{[18,20]}$&$\scriptstyle\langle {}^5A_{10,15} \left(\vec{e}_{10,15} - [ \scriptscriptstyle0 \; \scriptscriptstyle 3]^T\right) - \vec{1}_5 \cdot 0.3 \rangle $ \\ 
\midrule
\multicolumn{3}{c}{Return task}\\ 
\midrule
$\phi_1$ & $G_{[30,35]}$ & $\scriptstyle\langle {}^8A_i \left(\vec{x}_1 - [ \scriptscriptstyle0 \; \scriptscriptstyle0]^T\right) - \vec{1}_8 \cdot 0.2\rangle $ \\
$\phi_{1,i}$ &$G_{[30,40]}$ & $\scriptstyle\langle {}^8A_{1,i} \left(\vec{e}_{1,i} - [\scriptscriptstyle0 \; \scriptscriptstyle0]^T\right) - \vec{1}_8 \cdot 2 \rangle,\forall i= 2,\ldots15$\\
\bottomrule[1pt]
\end{tabular}
\caption{List of tasks for the exploration and return phase.}
\label{tab: tasks}
\end{table}

\begin{figure}
    \centering
         \includegraphics[scale=1.4]{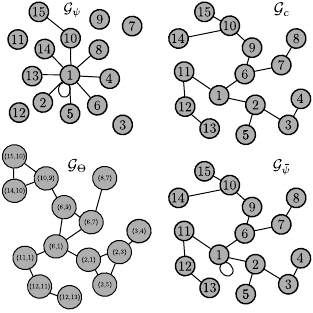}
        \caption{ (Upper-left) original task graph $\mathcal{G}_{\psi}$, (upper-right) new task graph $\mathcal{G}_{\bar{\psi}}$, (lower-left) edge-computing graph $\mathcal{G}_{\Theta}$ and (lower-right) communication graph $\mathcal{G}_c$,.}
    \label{fig: graphs}
\end{figure}

\begin{table*}[bt]
    \caption{Combined Edge-computing graph data}
    \label{tab:combined_summary_table}
    \centering
    \begin{tabular}{lccccccccccccccc}
        \toprule
        $(r,s)\in \mathcal{E}_{\pi}$ & \textbf{(10,9)} & \textbf{(15,10)} & \textbf{(2,1)} & \textbf{(4,3)} & \textbf{(14,10)} & \textbf{(1,11)} & \textbf{(13,12)} & \textbf{(12,11)} & \textbf{(6,1)} & \textbf{(5,2)} & \textbf{(9,6)} & \textbf{(8,7)} & \textbf{(3,2)} & \textbf{(7,6)} \\
        \midrule
        $|\Pi_{rs}|$                     & 4  & 1  & 5     & 2   & 2  & 3   & 2  & 3   & 8   & 2  & 5   & 2   & 3   & 3 \\
        \text{dim }$|\mathcal{X}_{rs}|$  & 14 & 3  & 21    & 6   & 6  & 13  & 6  & 11  & 30  & 6  & 17  & 6   & 11  & 11 \\
        $\sum_{(i,j)\in \Pi_{rs}}m_{ij}$ & 208 & 64 & 253  & 100 & 80 & 153 & 89 & 153 & 436 & 89 & 272 & 100 & 164 & 164 \\
        $|\mathcal{Q}_{rs}|$             & 1  & 0  & 3     & 0   & 0  & 2   & 0  & 1   & 3   & 0  & 1   & 0   & 1   & 1 \\
        \bottomrule
    \end{tabular}
\end{table*}
\begin{figure*}
    \centering
    \begin{subfigure}[t]{0.3\textwidth}
          \centering
         \includegraphics[height=6.2cm]{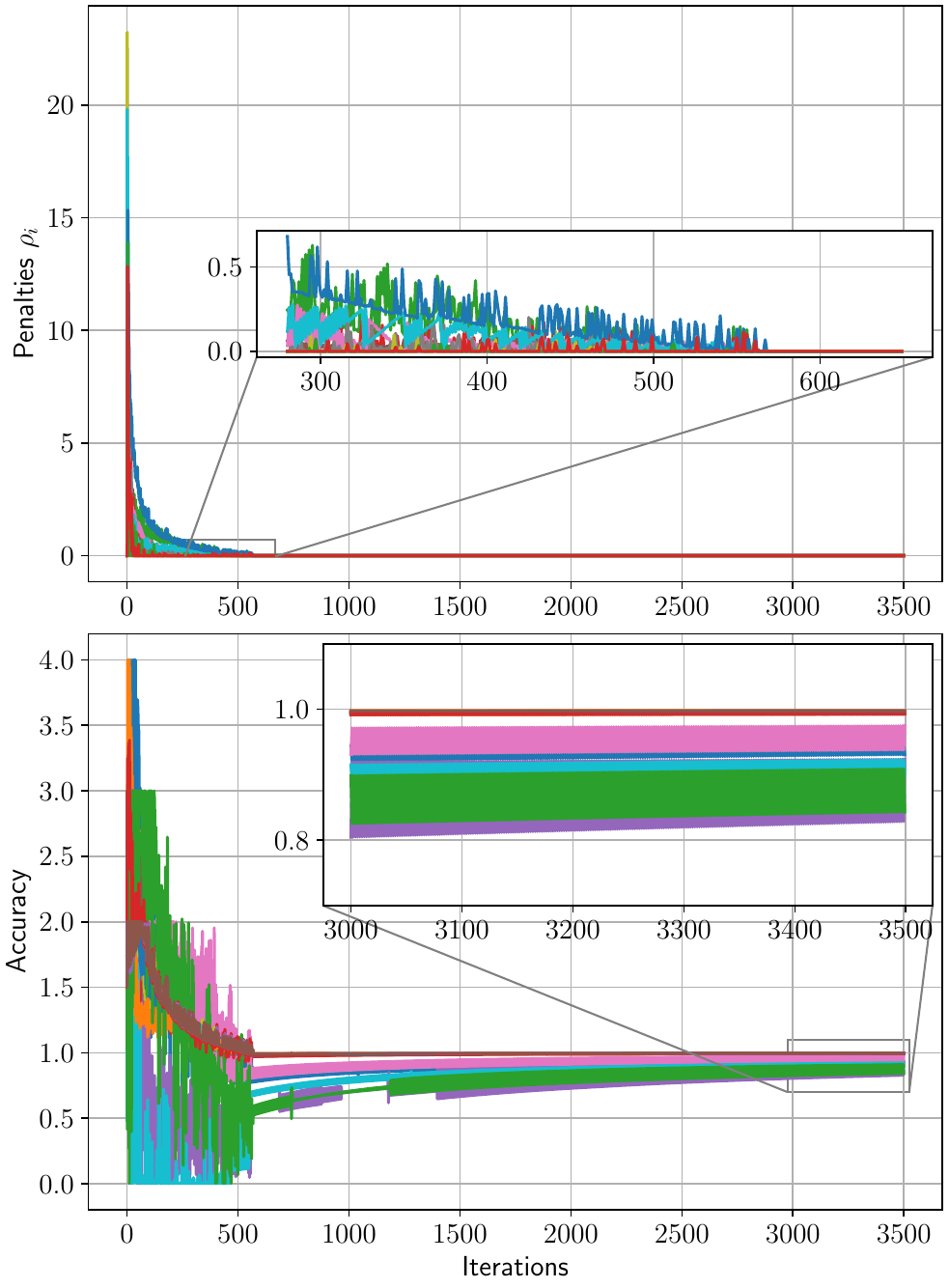}
        \caption{Evolution of penalties optimal values ${}^{t}\rho_{rs}\; \forall (r,s)\in \mathcal{E}_{\pi}$ as per Alg. \ref{alg:rsdd}}
                \label{fig:decomposition result}
    \end{subfigure}%
    \hfill
    \begin{subfigure}[t]{0.65\textwidth}
         \centering
         \includegraphics[height=6.2cm]{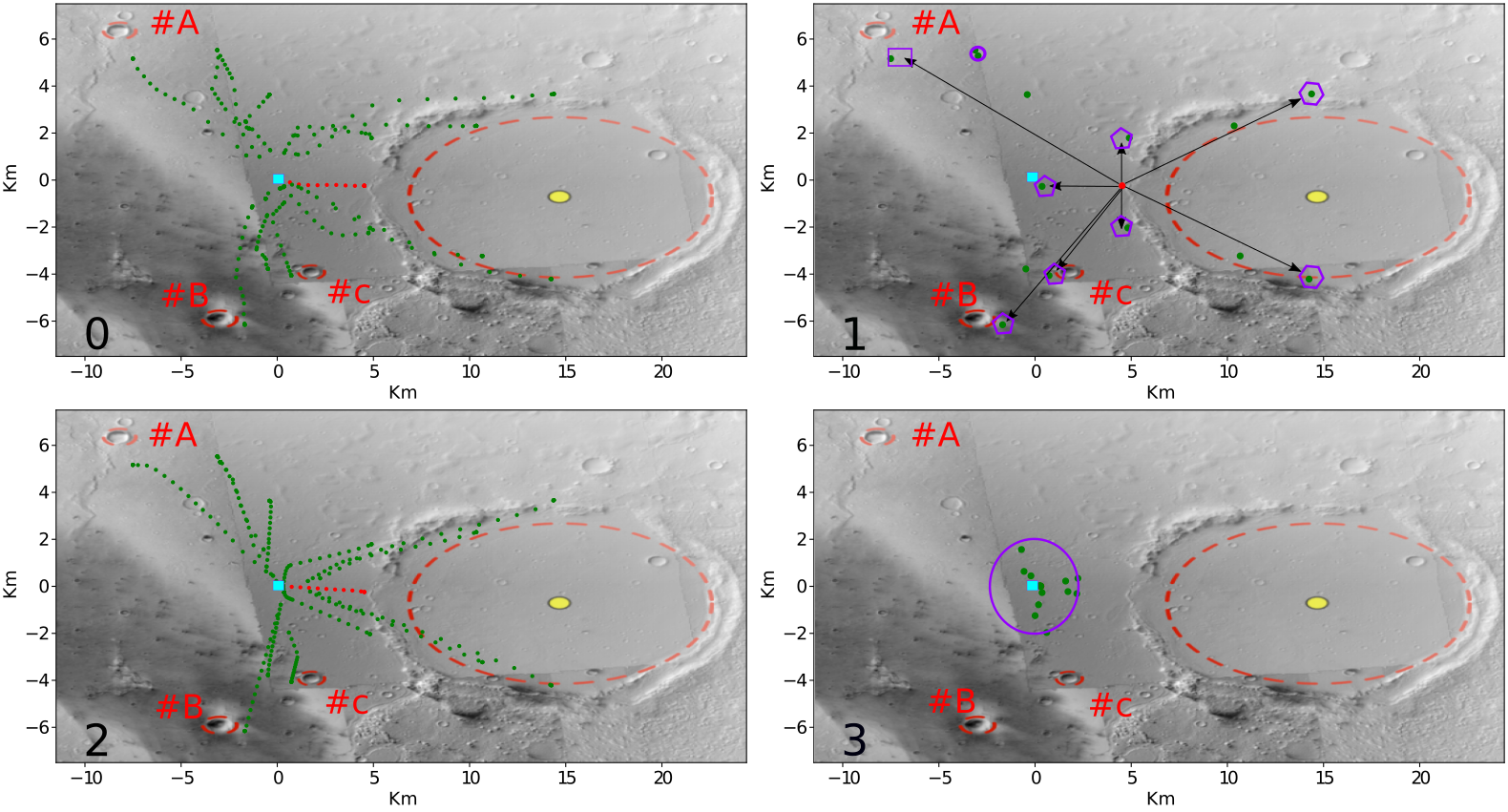}
        \caption{(Panel 0): Exploration phase over the interval 0-20h. (Panel 1): Snapshot of the achieved configuration at time 20h. (Panel 2) Return phase from time 20-40h. (Panel 3) Snapshot of the achieved configuration at time 40h.}
        \label{fig:mission panels}
    \end{subfigure}
    \caption{}
    \label{fig:simulation image}
\end{figure*}
We apply the developed algorithm for a space exploration task. A team of 15 agents with single integrator dynamics $\dot{\vec{x}}_i= \vec{u}_i  \in \mathbb{R}^2 , \forall i=1,\ldots 15$ is considered for the mission, where $\vec{x}_i$ represents the position of each agent and such that $\|\vec{u}_i\| \leq 1.8 \; km/h$. The exploration consists of taking three panoramic images of the Thira crater on Mars' southern hemisphere from three different sides of the crater and visiting smaller craters around Thira. The interest points are represented by red dashed circles in Figure \ref{fig:mission panels}, with Thira being the largest and marked by a yellow dot. A cyan square represents the location of the main base station from which the agents initially depart to accomplish the mission. Each agent is assumed to have a maximum communication range of $r_c =8.5$ km. The mission consists of a first exploration phase from the base station located at $[0,0]$ to the interest points, which lasts 20 hours. During the exploration phase, agent 1 can be considered the global leader of the team and the one holding the information about the interest points location such that the other agents 2-15 can reach their target location by staying in formation with agent 1 according to the collaborative tasks in Table \ref{tab: tasks}. Specifically, agents 1,8 and 4 take charge of taking the three panoramic images of Thira, with agent 1 (red trajectory in Fig. \ref{fig:mission panels}) approaching Thira from the west side leaving from the base station, while agents 4 and 8 reach the southern and northern sides of the crater respectively by maintaining an angle of 45 degrees with agents 1, thus reaching a triangular formation around Thira. Agent 14 visits the upper-left crater (marked as $\#A$), while agents 13 and 5 visit the lower craters ($\#B$ ad $\#C$ respectively). At the same time, agents 11,6 and 2 should achieve a triangular formation around 1. Lastly, agent 15 and 10 should remain within 200 m from each other. After the exploration phase, the agents should group again around agent 1 in a radius of 2 km, while agent 1 returns to the base station effectively ``dragging" the team back to base. The list of tasks for the exploration and return phase are shown in Table \ref{tab: tasks}, where $\mathcal{P}({}^{n}A_{ij}, \vec{c}_{ij},\vec{1}_n \beta_{ij})$ represents a regular polytope with $n$ sides, with each side having distance $\beta_{ij}>0$ from the centre $\vec{c}_{ij}$ (i.e $n=5$ represents a regular penthagon). It is not an assumption of our approach that the polytopes $\mathcal{P}$ should have any regularity. The agents are controlled via the Control Barrier Functions-based controller proposed in \cite{gregcdc}, such that each agent only exploits the state information of its communication neighbours in $\mathcal{N}_{c}$, where the communication graph is represented in Figure \ref{fig: graphs}. Figure \ref{fig:mission panels} shows the exploration phase on the two upper panels, where panel 0 represent the exploration from time 0h to 20h, while panel 1 represents a snapshot of the final configuration at time 20h. The return phase is represented in panels 2 and 3, where panel 2 shows the trajectories of the agents from time 20h to 40h, while panel 3 shows a snapshot of the final configuration at time 40h. Pink polytopes and black arrows apply to represent the original truth set of the tasks in Table \ref{tab: tasks}. The task graph ($\mathcal{G}_\psi$), communication graph ($\mathcal{G}_c$) and computing graph ($\mathcal{G}_{\Theta}$) are represented in Figure \ref{fig: graphs} together with the resulting task graph $\mathcal{G}_{\bar{\psi}}$ obtained implementing the distributed task decomposition algorithm presented in Section \ref{distributed task decompostion section}.  \par
The task decomposition is computed before the start of the mission at time 0h, where Alg \ref{alg:rsdd} is run for 3500 iterations by each agent in $\mathcal{G}_{\theta}$ with an average computational time of $52.133s$ (assuming no delay in exchanging the consensus and lagrangian multiplier variables). General information about the number of decomposition paths passing through a specific edge $(r,s)\in \mathcal{E}_{\pi}$ ($|\Pi_{rs}|$), the dimension of the variables $\vec{\chi}_{rs}$, the total number of shared constraints ($\sum_{(i,j) \in \Pi_{rs}}m_{ij}$)  and the number of conflicting conjunctions sets of tasks ($|\mathfrak{Q}_{rs}|$) are given in Table \ref{tab:combined_summary_table}. Figure \ref{fig:decomposition result} shows the evolution of the penalties values ${}^t\rho_{r,s} \; \forall (r,s)\in \mathcal{E}_{\pi}$ and the accuracy of decomposition for each task over a range of 3500 iterations. In particular, the time evolution of the optimal penalties ${}^{t}\rho_{rs}$ indicates that a valid task decomposition is found after 600 iterations of Alg. \ref{alg:rsdd}, while the decomposition accuracy as per \eqref{eq:task decomposition accuracy} keeps improving before stabilizing at 3500 iterations. We note that since the optimization program \eqref{eq:distributed opt} is only quasi-convex, the solution of the task decomposition is not unique. We used the open-source library CasADi (\cite{casadi}) to solve the optimization program \eqref{eq:distributed opt} leveraging an interior-point method solver on an Intel-Core i7-1265U. We highlight that agents in Figure \ref{fig:simulation image} only exploit information from neighbouring agents in $\mathcal{G}_c$ for both controlling their state in Figure \ref{fig:mission panels} and achieving the task decomposition before staring the mission.
\section{Conclusions}\label{conclusions}
We proposed a decentralized task decomposition approach that can be applied to decompose collaborative STL tasks, defined over communication inconsistent edges, as conjunctions of tasks defined over communication consistent edges. We provided a case study of a space exploration mission to corroborate the validity of the proposed approach. Future work is required to expand the proposed results to cyclic graphs, to consider more generic types of collaborative tasks and to include conflicting conjunctions among independent and collaborative tasks.


\bibliographystyle{ieeetr} 
\bibliography{references} 

\appendices
\section{} \label{appendixa}
\subsection{Proof of Proposition \ref{linear scaling and translation}}
\begin{proof}
    Let the generator set $ V=\{\vec{v}_k\}_{k=1}^{|V|} = \nu(\mathcal{P}(A,\vec{0},\vec{z}))$. We first prove that for all $\vec{x} \in \mathcal{P}(A,\vec{0},\vec{z})$ it holds $\vec{y}=\alpha \vec{x} + \vec{c} \in \mathcal{P}(A,\vec{c},\alpha\vec{z})$. First, note that by Prop. \ref{equivalent represenations} we have $\vec{x} = \sum_{k=1}^{|V|} \lambda_k\vec{v}_k, \forall \vec{x}\in \mathcal{P}(A,\vec{0},\vec{z})$ with $\sum_{k}^{|V|}\lambda_k=1$, $\lambda_k \geq 0$. Moreover, by definition of $\mathcal{P}(A,\vec{0},\vec{z})$ (cf. Def. \ref{h-polyedron}), all the vertices $\vec{v}_k \in  \nu(\mathcal{P}(A,\vec{0},\vec{z}))$ satisfy $A\vec{v}_k \leq \vec{z}$. Thus we have $A\vec{y} = A(\alpha \vec{x} + \vec{c}) = \alpha A(\sum_{k=1}^{|V|} \lambda_k\vec{v}_k + \vec{c}) \leq \alpha\sum_{k=1}^{|V|} \lambda_k \vec{z} + A\vec{c} \leq \alpha \vec{z}+ A\vec{c}$. Thus we just proved that $A\vec{y}(\vec{x}) \leq \vec{z}+ A\vec{c},\; \forall \vec{x} \in \mathcal{P}(A,\vec{0},\vec{z}) $ from which we have $\mathcal{P}(A,\vec{c},\alpha\vec{z}) = \{ \vec{y} | A(\vec{y}-\vec{c}) \leq \vec{z}\} = \{ \vec{c} + \alpha \vec{x} | \vec{x} \in \mathcal{P}(A,\vec{0},\vec{z})\}$. We then conclude noting that every $\vec{y} \in \mathcal{P}(A,\vec{c},\alpha\vec{z})$ is uniquely defined by convex combinations of the generators $\{\alpha\vec{v}_k + \vec{c}\}_{k=1}^{|V|}$, which can be compactly written as $\{G_k \begin{bmatrix}
        \vec{c}\\ \alpha
    \end{bmatrix}\}_{k =1,\ldots |V|}$ with $G_k=\comp{I_n}{\vec{v}_k}$.
\end{proof}

\subsection{Proof of Proposition \ref{polytopes inclusion}}
\begin{proof}
For both \eqref{eq:compact inclusion} and \eqref{eq:compact intersection} we derive the reverse equivalence ($\Leftarrow$) while the forward equivalence ($\Rightarrow$) can be obtained trivially. We first prove the inclusion relation \eqref{eq:compact inclusion}. Namely, consider the set of generators $V_1 =  \nu(\mathcal{P}(A_1,\vec{0},\vec{z}_1))$ such that by Prop. \eqref{eq:generators equivalence} we have $\nu(\mathcal{P}(A_1,\vec{c}_1,\alpha_1\vec{z}_1)) = \{G_k \vec{\eta}_1\}_{k=1}^{|V_1|}$ with $G_k = \comp{[I_n}{\vec{v}_k},\; \forall \vec{v}_k \in V_1$. Then the inclusion of every generator vector in $\{G_k \vec{\eta}_1\}_{k=1}^{|V_1|}$, in the set $\mathcal{P}(A_2,\vec{c}_2,\alpha_2\vec{z}_2))$ is expressed as 
\begin{equation}\label{eq:first relation}
A_2(G_k \vec{\eta}_1 -\vec{c}_2)-\alpha_2 \vec{z}_2 \leq \vec{0},\; \forall k=1,\ldots |V_1|,
\end{equation}
which after rearrangement of the terms becomes
\begin{equation}\label{eq:second relation}
A_2G_k \vec{\eta}_1 - \comp{A_2}{\vec{z}_2}\vec{\eta}_2  \leq \vec{0}\; \forall k=1,\ldots |V_1|.
\end{equation}
Notice now that \eqref{eq:compact inclusion} is a compact matrix representation of this last set of inequalities. Since $\mathcal{P}(A_1,\vec{c}_1,\alpha_1\vec{z}_1)$ can be described by the convex hull of its generators in $\{G_k \vec{\eta}_1\}_{k=1}^{|V_1|}$ as per Def \ref{convex hull}, we have that every vector $\vec{x}\in \mathcal{P}(A_1,\vec{c}_1,\alpha_1\vec{z}_1)$ can be described as $\vec{x} = \sum_k\lambda_k G_k\vec{\eta}_1$ with  $\lambda_k \geq 1$ and $\sum_k \lambda_k = 1$ such that by \eqref{eq:first relation} it holds
$$
A_2(\vec{x} -\vec{c}_2) - \alpha_2\vec{z}_2 = A_2(\sum_k\lambda_kG_k\vec{\eta_1} -\vec{c}_2) - \alpha_2\vec{z}_2 \leq \vec{0}.
$$
This last relation then proves $\vec{x}\in \mathcal{P}(A_2,\vec{c}_2,\vec{z}_{2})$ for all $\vec{x}\in \mathcal{P}(A_1,\vec{c}_1,\vec{z}_{1})$, thus proving the inclusion. Turning to the intersection relation  \eqref{eq:compact intersection},  we have that if there exists a vector $\vec{\xi}$ such that  $A_1(\vec{\xi}-\vec{c}_1) - \alpha_1 \vec{z}_1 \leq \vec{0} \land  A_1(\vec{\xi}-\vec{c}_1) - \alpha_1 \vec{z}_1 \leq \vec{0}$, then, by definition of polytope, it holds $\vec{\xi}\in \mathcal{P}(A_2,\vec{c}_2,\vec{z}_2)$ and $\vec{\xi}\in \mathcal{P}(A_1,\vec{c}_1,\vec{z}_1)$. Thus necessarily $\mathcal{P}(A_1,\vec{c}_1,\vec{z}_1) \cap \mathcal{P}(A_2,\vec{c}_2,\vec{z}_2)\supseteq \{\vec{\xi}\} \neq \emptyset$, concluding the proof.
\end{proof}

\subsection{Proof of Fact \ref{conflict 1}}
\begin{proof}
We prove the fact by contradiction. Consider there exists $L\in \mathfrak{L}_{ij}$ with $\bigcap_{l \in L} [a^{l},b^{l}] \neq \emptyset$,  $\bigcap_{l \in L} \mathcal{B}_{ij}^{l}=\emptyset$ and let $\vec{x}(t)$ be a signal such that $(\vec{x}(t),0)\models \bigwedge_{k=1}^{K_{ij}}\varphi^{k}_{ij}$. By definition of the robust semantics  \eqref{eq:robust semantics}, the satisfaction of  $\bigwedge_{k=1}^{K_{ij}}\varphi^{k}_{ij}$ implies (omitting the argument $(\vec{x}(t),0)$)
$$
\min \{\min_{l\in L} \{\rho^{\varphi^l_{ij}}\}, \min_{l\in \mathcal{I}_G\setminus L} \{\rho^{\varphi^l_{ij}}\}, \min_{d\in \mathcal{I}_F} \{\rho^{\varphi^d_{ij}}\}  \}\geq 0,
$$
and thus $\min_{l\in L} \{\rho^{\varphi^l_{ij}}\} \geq 0$. By definition we have that $\min_{l\in L} \{\rho^{\varphi^l_{ij}}\} \geq 0 \Leftrightarrow h^l_{ij}(\vec{e}_{ij}(t)) \geq 0,\;  \forall t\in [a^l,b^l],\; \forall l\in L \Leftrightarrow \vec{e}_{ij}(t) \in \mathcal{B}_{ij}^l, \; \forall t\in [a^l,b^l],\; \forall l\in L$, where the first equivalence derives from the definition of robust semantics for the operator $G$ in \eqref{eq:always robust}, while the second implication derives from the definition of truth set \eqref{eq:super level sets}. At the same time, we have initially assumed that $\bigcap_{l \in L} [a^{l},b^{l}] \neq \emptyset$ such that for all $t \in \bigcap_{l \in L} [a^{l},b^{l}]$ it must hold $\vec{e}_{ij}(t) \in \mathcal{B}_{ij}^l\; \forall l\in L \Leftrightarrow \vec{e}_{ij}(t) \in \bigcap_{l \in L} \mathcal{B}_{ij}^{l}$. We thus arrived at the contradiction since we initially assumed that $\bigcap_{l \in L} \mathcal{B}_{ij}^{l}=\emptyset$.
\end{proof}

\subsection{Proof of Fact \ref{conflict 2}}
\begin{proof}
We prove the result by contradiction.  Consider there exists an index $\bar{d}\in \mathcal{I}_{F}$ with task $\phi_{ij}^{\bar{d}}$ and let $C$ be a subset of indices in $\mathcal{I}_G$ such that $C \in \mathfrak{C}_{ij}(\bar{d})$ and $\mathcal{B}_{ij}^{\bar{d}}\cap\mathcal{B}_{ij}^{l}=\emptyset, \; \forall l\in C$. Moreover, assume there exists a signal $\vec{x}(t)$ such that $(\vec{x}(t),0)\models \bigwedge_{k=1}^{K_{ij}}\varphi^{k}_{ij}$. By definition of the robust semantics  \eqref{eq:robust semantics}, the satisfaction of  $\bigwedge_{k=1}^{K_{ij}}\varphi^{k}_{ij}$ implies (omitting the argument $(\vec{x}(t),0)$)
$$
\min \{\min_{l\in C} \{\rho^{\varphi^l_{ij}}\}, \min_{l\in \mathcal{I}_G\setminus C} \{\rho^{\varphi^l_{ij}}\},\rho^{\varphi_{ij}^{\bar{d}}}, \min_{d\in \mathcal{I}_F\setminus \{\bar{d}\}} \{\rho^{\varphi^d_{ij}}\}  \}\geq 0,
$$
and thus $\rho^{\varphi_{ij}^{\bar{d}}}\geq 0$, $\min_{l\in C} \{\rho^{\varphi^l_{ij}}\}\geq 0$. Now by the robust semantics in \eqref{eq:always robust}, \eqref{eq:conjunction robust} we have that $\min_{l\in C} \{\rho^{\varphi^l_{ij}}\}\geq 0$ implies that for all $t\in \cup_{l\in C}[a^l,b^l]$ it holds $\bigvee_{l\in C} h_{ij}^{l}(\vec{e}_{ij}(t))\geq 0 \Leftrightarrow  \bigvee_{l\in C} \vec{e}_{ij}(t) \in \mathcal{B}_{ij}^l$, where \eqref{eq:super level sets} applies for the last equivalence. In other words, for at least one $l\in C$ it must hold $\vec{e}_{ij}(t) \in \mathcal{B}_{ij}^l$ at every time $ t\in \cup_{l\in C}[a^l,b^l]$. On the other hand, by the robust semantics for the \textit{F} operator in \eqref{eq:eventually robust}, it holds $\rho^{\varphi_{ij}^{\bar{d}}}\geq 0 \Leftrightarrow h_{ij}^{\bar{d}}(\vec{e}_{ij}(\tau))\geq 0 \Leftrightarrow \vec{e}_{ij}(\tau) \in \mathcal{B}_{ij}^{\bar{d}}$ for some  $\tau \in [a^{\bar{d}},b^{\bar{d}}]$. But since $[a^{\bar{d}},b^{\bar{d}}]\subseteq \cup_{l\in C}[a^l,b^l]$ then for any $\tau \in [a^{\bar{d}}, b^{\bar{d}}] \subseteq \cup_{l\in C}[a^l,b^l]$ there must exists at least one $l \in C$ such that $\vec{e}_{ij}(\tau) \in \mathcal{B}_{ij}^{l} \cap \mathcal{B}_{ij}^{\bar{d}}$. We thus arrived at the contradiction since we initially assumed $\mathcal{B}_{ij}^{l} \cap \mathcal{B}_{ij}^{\bar{d}} = \emptyset, \forall l\in C$. 
\end{proof}
\subsection{Proof of Fact \ref{conflict 3}}
\begin{proof}
We prove the result by contradiction. Consider there exists an index $\bar{d}\in \mathcal{I}_{F}$ with task $\phi_{ij}^{\bar{d}}$ and a set $D \in \mathfrak{D}_{ij}(\bar{d})$ with $\mathcal{B}_{ij}^{\bar{d}} \cap \bigcap_{l\in D} \mathcal{B}_{ij}^{l}=\emptyset$. Leveraging the definition of robust semantics we have by \eqref{eq:robust semantics}
$$
\min \{\min_{l\in D} \{\rho^{\varphi^l_{ij}}\}, \min_{l\in \mathcal{I}_G\setminus D} \{\rho^{\varphi^l_{ij}}\},\rho^{\varphi_{ij}^{\bar{d}}}, \min_{d\in \mathcal{I}_F\setminus \{\bar{d}\}} \{\rho^{\varphi^d_{ij}}\}  \}\geq 0 ,
$$
and thus $\min_{l\in D} \{\rho^{\varphi^l_{ij}}\}\geq 0$, $\rho^{\varphi_{ij}^{\bar{d}}}\geq 0$. By the same proof of Fact 1, we know that $\min_{l\in D} \{\rho^{\varphi^l_{ij}}\}\geq 0 \Leftrightarrow \vec{e}_{ij}(t)\in \bigcap_{l\in D} \mathcal{B}_{ij}^l,\; \forall t\in \bigcap_{l\in D}[a^l,b^l]$. At the same time, by the definition of the robust semantics for the operator $F$ in \eqref{eq:eventually robust} we have that  $\rho^{\varphi_{ij}^{\bar{d}}}\geq 0 \Leftrightarrow \exists \tau \in [a^{\bar{d}}, b^{\bar{b}}], \text{s.t} \; h_{ij}^{\bar{d}}(\tau)\geq 0$ and, in turn, $\vec{e}_{ij}(\tau) \in \mathcal{B}_{ij}^{\bar{d}}$ by \eqref{eq:super level sets}. On the other hand, by definition of the set $\mathfrak{D}_{ij}(\bar{d}) \ni D$ we have that $[a^{\bar{d}}, b^{\bar{d}}] \subseteq \bigcap_{l\in D}[a^l,b^l]$ such that $\tau \in \bigcap_{l\in D}[a^l,b^l]$. Thus there must exist a $\tau \in [a^{\bar{d}}, b^{\bar{d}}]$ such that $\vec{e}_{ij}(\tau) \in \mathcal{B}_{ij}^{\bar{d}}$, but also $\vec{e}_{ij}(t)\in \bigcap_{l\in D} \mathcal{B}_{ij}^l,\; \forall t \in  [a^{\bar{d}}, b^{\bar{d}}]$. Then, necessarily  $\vec{e}_{ij}(\tau) \in \mathcal{B}_{ij}^{\bar{d}} \cap  \bigcap_{l\in D} \mathcal{B}_{ij}^l \neq \emptyset$ contradicting the initial assumption.
\end{proof}

\subsection{Proof  Fact \ref{conflict 4}} 
\begin{proof}
We prove the fact by contradiction. Assume there exists a signal $\vec{x}(t)$ such that $(\vec{x}(t),0) \models \land_{(r,s) \in \epsilon(\vec{\omega})} \varphi_{rs}$ and $\vec{0} \not\in \bigoplus_{(r,s) \in \epsilon(\vec{\omega})}\mathcal{B}_{rs}$. Then from the robust semantics \eqref{eq:conjunction robust},\eqref{eq:always robust} and from \eqref{eq:super level sets}, it holds $\rho^{\land_{(r,s) \in \epsilon(\vec{\omega})} \varphi_{rs}}(\vec{x}(t),0) = \min_{(r,s)\in \epsilon(\vec{\omega})} \{ \rho^{\varphi_{rs}}(\vec{x}(t),0) \} \geq 0 \Rightarrow \rho^{\varphi_{rs}}(\vec{x}(t),0) \geq 0, \; \forall (r,s)\in \epsilon(\vec{\omega}) \Rightarrow h_{rs}(\vec{e}_{rs}(t))\geq0,  
\forall t \in [a_{rs},b_{rs}],\;\forall (r,s)\in \epsilon(\vec{\omega}) \Rightarrow \vec{e}_{rs}(t) \in \mathcal{B}_{rs} ,\; \forall t \in [a_{rs},b_{rs}],\;\forall (r,s)\in \epsilon(\vec{\omega})$. Since, by assumption, $\bigcap_{(r,s)\in \epsilon(\vec{\omega})}[a_{rs},b_{rs}]\neq \emptyset$, then there exist a time instant $\tau \in \bigcap_{(r,s)\in \epsilon(\vec{\omega})}[a_{rs},b_{rs}]$ such that  $\vec{e}_{rs}(\tau) \in \mathcal{B}_{rs},\;\forall (r,s)\in \epsilon(\vec{\omega})$ and thus $\sum_{(r,s)\in \epsilon(\vec{\omega})}\vec{e}_{rs}(\tau) \in \bigoplus_{(r,s)\in \epsilon(\vec{\omega})} \mathcal{B}_{rs}$. By \eqref{eq:edge cycle} it also holds $\sum_{(r,s)\in \epsilon(\vec{\omega})}\vec{e}_{rs}(\tau) = \vec{0}$. We thus arrived at the contradiction since these last two conditions only hold if $\vec{0} \in \bigoplus_{(r,s) \in \epsilon(\omega)}\mathcal{B}_{rs}$.
\end{proof}
\par
\subsection{Proof  Fact \ref{conflict 5}}
\begin{proof}
The proof is similar to the proof of Fact \ref{conflict 4} in the sense that there must exists a time $\tau \in [a_{ip},b_{ip}] \subseteq \cap_{(r,s)\in \epsilon(\vec{\pi}_p^i)}[a_{rs},b_{rs}]$ such that  $\vec{e}_{rs}(\tau) \in \mathcal{B}_{rs},\; \forall (r,s) \in \epsilon(\vec{\omega})$ to satisfy the conjunction of tasks $\land_{(r,s) \in \epsilon(\vec{\omega})} \varphi_{rs}$. At the same time, the cycle closure condition \eqref{eq:edge cycle} must hold at time $\tau$. The rest of the proof follows the proof of Fact \ref{conflict 4}.
\end{proof}
\end{document}